\documentclass[12pt, draftclsnofoot, peerreview, letter, oneside, onecolumn]{IEEEtran}
\usepackage{graphicx}
\usepackage{amsmath}
\usepackage{amssymb}
\usepackage{mathrsfs}
\usepackage{stfloats}
\usepackage{dsfont}
\usepackage{setspace}
\usepackage{array}
\usepackage{epstopdf}
\usepackage{caption}
\usepackage[font={footnotesize}]{caption}
\newtheorem{corollary}{Corollary}
\newtheorem{proposition}{Proposition}
\newtheorem{lemma}{Lemma}
\newtheorem{remark}{Remark}
\begin{document}
%
%
%
%
%
%
\title{Stochastic Geometry Modeling and Analysis of Multi-Tier Millimeter Wave Cellular Networks \vspace{-0.25cm}}
\author{Marco~Di~Renzo,~\IEEEmembership{Senior~Member,~IEEE}
\thanks{Manuscript received October 14, 2014. M. Di Renzo is with CNRS--SUPELEC--University of Paris--Sud XI, 3 rue Joliot--Curie, 91192 Gif--sur--Yvette, France (e--mail: marco.direnzo@lss.supelec.fr). This work is supported in part by the European Commission under the auspices of the FP7--PEOPLE MITN--CROSSFIRE project (grant 317126).} }
%
%
\markboth{Transactions on Wireless Communications} {M. Di Renzo: Stochastic Geometry Modeling and Analysis of Multi-Tier Millimeter Wave Cellular Networks}
\maketitle
\vspace{-0.75cm}
\begin{abstract}
\vspace{-0.45cm}
In this paper, a new mathematical framework to the analysis of millimeter wave cellular networks is introduced. Its peculiarity lies in considering realistic path-loss and blockage models, which are derived from recently reported experimental data. The path-loss model accounts for different distributions of line-of-sight and non-line-of-sight propagation conditions and the blockage model includes an outage state that provides a better representation of the outage possibilities of millimeter wave communications. By modeling the locations of the base stations as points of a Poisson point process and by relying on a noise-limited approximation for typical millimeter wave network deployments, simple and exact integral as well as approximated and closed-form formulas for computing the coverage probability and the average rate are obtained. With the aid of Monte Carlo simulations, the noise-limited approximation is shown to be sufficiently accurate for typical network densities. The proposed mathematical framework is applicable to cell association criteria based on the smallest path-loss and on the highest received power. It accounts for beamforming alignment errors and for multi-tier cellular network deployments. Numerical results confirm that sufficiently dense millimeter wave cellular networks are capable of outperforming micro wave cellular networks, both in terms of coverage probability and average rate.
\end{abstract}
\vspace{-0.5cm}
\begin{keywords}
\vspace{-0.45cm}
Millimeter Wave Communications, Multi-Tier Cellular Networks, Stochastic Geometry.
\end{keywords}
\vspace{-0.25cm}
\section{Introduction} \label{Introduction}
\vspace{-0.25cm}
\PARstart{I}{n} spite of common belief, recently conducted channel measurements have shown that millimeter wave (mmWave) frequencies may be suitable for cellular communications, provided that the cell radius is of the order of 100-200 meters \cite{Rappaport__mmWaveAccess}. Based on these measurements, the authors of \cite{Rappaport__mmWaveJSAC} have recently investigated system-level performance of mmWave cellular networks and have compared them against conventional micro wave ($\mu$Wave) cellular networks. The obtained results have highlighted that mmWave cellular communications may outperform $\mu$Wave cellular communications, by assuming similar cellular network densities, provided that a sufficient beamforming gain is guaranteed between Base Stations (BSs) and Mobile Terminals (MTs). These preliminary but encouraging results have motivated several researchers to investigate potential and challenges of mmWave cellular communications for wireless access, in light of the large and unused spectrum that is available at these frequencies \cite{Rangan}, \cite{Ghosh}.

System-level performance evaluation of cellular networks is widely recognized to be a mathematically intractable problem \cite{AndrewsNov2011}. This is due to the lack of tractable approaches for modeling the locations of the BSs and the other-cell interference. Only recently, a new mathematical methodology has gained prominence due to its analytical tractability, its capability of capturing the inherent performance trends of currently deployed cellular networks, and the possibility of studying next-generation heterogeneous network deployments. This emerging approach exploits results from stochastic geometry and relies on modeling the locations of the BSs as points of a point process \cite{AndrewsNov2011}. Usually, the Poisson Point Process (PPP) is used due to its mathematical tractability \cite{Haenggi_Survey2013}. Recent results on cellular networks modeling based on stochastic geometry are available in \cite{MDR_TCOMrate}-\cite{MDR_COMMLPeng}, to which the reader is referred for a comprehensive literature review.

Motivated by the mathematical flexibility of the PPP-based abstraction modeling, researchers have recently turned their attention to study system-level performance of mmWave cellular networks with the aid of stochastic geometry. The aim is to develop mathematical frameworks specifically tailored to account for the peculiarities of mmWave propagation channels and transmission schemes \cite{Heath__CommunMag}-\cite{Singh__mmWave}. In fact, currently available mathematical frameworks for modeling $\mu$Wave cellular networks are not directly applicable to mmWave cellular networks. The main reasons are related to the need of incorporating realistic path-loss and blockage models, which are significantly different from $\mu$Wave communications. For example, the authors of \cite{Rappaport__mmWaveAccess} and \cite{Rappaport__mmWaveJSAC} have pointed out that Line-Of-Sight (LOS) and Non-Line-Of-Sight (NLOS) links need to be appropriately modeled and may have different distributions, due to the more prominent impact of spatial blockages at mmWave frequencies compared to $\mu$Wave frequencies. Also, in mmWave communications a new outage state may be present in addition to LOS and NLOS states, which better reflects blockage effects at high frequencies and accounts for the fact that a link may be too weak to be established. In addition, large-scale antenna arrays are expected to be used for directional beamforming in mmWave systems, in order to overcome the increased path-loss at mmWave frequencies and to provide other-cell interference isolation. Therefore, directional beamforming needs to be included in the mathematical modeling and performance evaluation.

Recently reported results on stochastic geometry modeling of mmWave cellular communications take these aspects into account only in part \cite{Heath__mmWave}, \cite{Singh__mmWave}. The approach proposed in \cite{Heath__mmWave} is mathematically tractable for dense cellular network deployments and relies on an equivalent LOS ball approximation. The interference-limited nature of the mmWave cellular networks analyzed in \cite{Heath__mmWave} is, in part, due to the considered ultra-dense network deployment and to the relatively small transmission bandwidth used for the analysis (\textit{i.e.}, 100 MHz). Larger transmission bandwidths of the order of 1-2 GHz are, on the other hand, expected to be used in future mmWave cellular systems \cite{Rappaport__mmWaveJSAC}, \cite{Ghosh}, \cite{Singh__mmWave}, \cite{PIMRC_2014}, which may enhance the impact of thermal noise compared to the other-cell interference. The approach proposed in \cite{Singh__mmWave} uses a similar LOS ball approximation, but is applicable to medium/sparse network deployments. Furthermore, it is validated by using actual building locations from dense urban deployments in the cities of New York and Chicago. The three-state link statistical model empirically derived in \cite{Rappaport__mmWaveJSAC} is, however, not explicitly taken into account either in \cite{Heath__mmWave} or in \cite{Singh__mmWave}. Also, the impact of cell association criteria, beamforming alignment errors and multi-tier other-cell interference are not considered in \cite{Heath__mmWave} and \cite{Singh__mmWave}. Similar comments apply to \cite{Singh__Globe2014}, which highlights the importance of considering realistic blockage models for accurate system-level performance evaluation of mmWave cellular communications. Against this background, in the present paper a new methodology to the stochastic geometry modeling and performance evaluation of mmWave cellular communications is proposed, which explicitly accounts for the empirical path-loss and blockage models reported in \cite{Rappaport__mmWaveJSAC}, for different cell association criteria, beamforming pointing errors and multi-tier deployments.

The paper is organized as follows. In Section \ref{SystemModel}, system model and modeling assumptions are introduced. In Section \ref{PPP_PathLoss}, the statistical distribution of deterministic and random transformations of the path-loss is provided. In Section \ref{CoverageRate}, the frameworks for computing coverage and rate of cellular networks are described, by considering a cell association based on the smallest path-loss and on the highest received power. In Section \ref{Generalizations}, the frameworks in Section \ref{CoverageRate} are generalized, by incorporating beamforming alignment errors and multi-tier deployments. In Section \ref{Results}, the analysis is validated via numerical simulations and the performance of mmWave and $\mu$Wave cellular networks are compared. Finally, Section \ref{Conclusion} concludes this paper.
\vspace{-0.1cm}
\section{System Model} \label{SystemModel}
\vspace{-0.1cm}
\subsection{PPP-Based Abstraction Modeling} \label{PPP_CellularModeling}
A bi-dimensional downlink cellular network is considered, where a probe MT is located, without loss of generality thanks to the Slivnyak theorem \cite[vol. 1, Th. 1.4.5]{BaccelliBook2009}, at the origin and the BSs are modeled as points of a homogeneous PPP, denoted by $\Psi $, of density $\lambda$. The MT is assumed to be served by the BS providing either the smallest path-loss (Section \ref{CoverageRate_PathLoss}) or the highest received power (Section \ref{CoverageRate_Best}) to it. The serving BS is denoted by ${\rm{BS}}^{\left( 0 \right)}$. Similar to \cite[Sec. VI]{AndrewsNov2011}, full-frequency reuse is considered. For notational simplicity, the set of interfering BSs is denoted by $\Psi ^{\left( \backslash 0 \right)}  = \Psi \backslash {\rm{BS}}^{\left( 0 \right)}$. The distance from a generic BS to the MT is denoted by $r$.
\subsection{Directional Beamforming Modeling} \label{BeamformingModeling}
Thanks to the small wavelength, mmWave cellular networks are capable of exploiting directional beamforming for compensating for the increased path-loss at mmWave frequencies and for overcoming the additional noise due to the large transmission bandwidth. As a desirable bonus, directional beamforming provides interference isolation, which reduces the impact of the other-cell interference. Thus, antenna arrays are assumed at both the BSs and the MT for performing directional beamforming. For mathematical tractability and similar to \cite{Heath__mmWave}, \cite{Singh__mmWave}, the actual antenna array patterns are approximated by a sectored antenna model. In particular, the antenna gain of a generic BS, $G_{{\rm{BS}}} \left(  \cdot  \right)$, and of the MT, $G_{{\rm{MT}}} \left(  \cdot  \right)$, can be formulated as follows:
\begin{equation}
\label{Eq_1}
G_{\rm{q}} \left( \theta  \right) = \begin{cases}
 G_{\rm{q}}^{\left( {\max } \right)} & {\rm{if}}\quad \left| \theta  \right| \le \omega _{\rm{q}}  \\
 G_{\rm{q}}^{\left( {\min } \right)} & {\rm{if}}\quad \left| \theta  \right| > \omega _{\rm{q}}  \\
 \end{cases}
\end{equation}
\noindent where $q \in \left\{ {{\rm{BS}},{\rm{MT}}} \right\}$, $\theta  \in \left[ {0,2\pi } \right)$ is the angle off the boresight direction, $\omega _q $ is the beamwidth of the main lobe, $G_q^{\left( {\max } \right)}$ and $G_q^{\left( {\min } \right)}$ are the array gains of main and side lobes, respectively.

The MT and its serving BS, ${\rm{BS}}^{\left( 0 \right)}$, are assumed to estimate the angles of arrival and to adjust their antenna steering orientations accordingly. In the absence of alignment errors, therefore, the maximum directivity gain can be exploited on the intended link. Thus, the directivity gain of the intended link is $G^{\left( 0 \right)}  = G_{{\rm{BS}}}^{\left( {\max } \right)} G_{{\rm{MT}}}^{\left( {\max } \right)}$. The beams of all non-intended links are assumed to be randomly oriented with respect to each other and to be uniformly distributed in $\left[ {0,2\pi } \right)$. Accordingly, the directivity gains of the interfering links, $G^{\left( i \right)}$ for $i \in \Psi ^{\left(\backslash 0 \right)}$, are randomly distributed. Based on \eqref{Eq_1}, their Probability Density Function (PDF) can be formulated as follows:
\begin{equation}
\label{Eq_2}
f_{G^{\left( i \right)} } \left( g \right) = \frac{{\omega _{\rm{q}} }}{{2\pi }}\delta \left( {g - G_q^{\left( {\max } \right)} } \right) + \left( {1 - \frac{{\omega _{\rm{q}} }}{{2\pi }}} \right)\delta \left( {g - G_q^{\left( {\min } \right)} } \right)
\end{equation}
\noindent where $q \in \left\{ {{\rm{BS}},{\rm{MT}}} \right\}$ and $\delta \left(  \cdot  \right)$ is the Kronecker's delta function.
\subsection{Beamforming Alignment Errors Modeling} \label{PointingErrorModeling}
The maximum directivity gain offered by directional beamforming, \textit{i.e.}, $G^{\left( 0 \right)}  = G_{{\rm{BS}}}^{\left( {\max } \right)} G_{{\rm{MT}}}^{\left( {\max } \right)}$, can be achieved only in the absence of beamsteering errors. Due to practical considerations, however, perfectly aligning the transmitter and the receiver may be difficult, especially if the beamwidth of the main lobe is quite small, as in mmWave systems for enhancing other-cell interference isolation. Indeed, the authors of \cite{Weber__PointingErrors} have recently reported that several tradeoffs affecting the performance of directional networks emerge in the presence of beamforming alignment errors. In the present paper, a beamsteering error model similar to \cite{Weber__PointingErrors} is considered.

Let $\theta _q^*$ for $q \in \left\{ {{\rm{BS}},{\rm{MT}}} \right\}$ be the angles corresponding to error-free beamsteering. Let $\varepsilon _q$ for $q \in \left\{ {{\rm{BS}},{\rm{MT}}} \right\}$ denote the additive beamsteering errors. In particular, $\varepsilon _{{\rm{BS}}}$ and $\varepsilon _{{\rm{MT}}}$ are assumed to be randomly distributed, to be independent of each other and to have a symmetric distribution around $\theta _{{\rm{BS}}}^*$ and $\theta _{{\rm{MT}}}^*$, respectively. Let $F_{\left| {\varepsilon _{\rm{q}} } \right|} \left( x \right) = \Pr \left\{ {\left| {\varepsilon _{\rm{q}} } \right| \le x} \right\}$ be the Cumulative Distribution Function (CDF) of ${\left| {\varepsilon _{\rm{q}} } \right|}$ for $q \in \left\{ {{\rm{BS}},{\rm{MT}}} \right\}$. Then, the PDF of the directivity gain of the intended link, $G^{\left( 0 \right)}$, can be formulated as $f_{G^{\left( 0 \right)} } \left( g \right) = \left( {f_{G_{{\rm{BS}}} }  \otimes f_{G_{{\rm{MT}}} } } \right)\left( g \right)$, where $ \otimes$ denotes the convolution operator, and $f_{G_{{\rm{BS}} } } \left(  \cdot  \right)$ and $f_{G_{{\rm{MT}}} } \left(  \cdot  \right)$ are the PDFs of the directivity gains of serving BS, ${\rm{BS}}^{\left( 0 \right)}$, and MT, which, from \eqref{Eq_1}, can be explicitly written as follows ($q \in \left\{ {{\rm{BS}},{\rm{MT}}} \right\}$):
\begin{equation}
\label{Eq_3}
f_{G_q } \left( g \right) = F_{\left| {\varepsilon _{\rm{q}} } \right|} \left( {\frac{{\omega _q }}{2}} \right)\delta \left( {g - G_q^{\left( {\max } \right)} } \right) + \left( {1 - F_{\left| {\varepsilon _{\rm{q}} } \right|} \left( {\frac{{\omega _q }}{2}} \right)} \right)\delta \left( {g - G_q^{\left( {\min } \right)} } \right)
\end{equation}

For example, if the beamsteering errors follow a Gaussian distribution with mean equal to zero and variance equal to $\sigma _{q,{\rm{BE}}}^2$ for $q \in \left\{ {{\rm{BS}},{\rm{MT}}} \right\}$, then ${\left| {\varepsilon _{\rm{q}} } \right|}$ follows a half-normal distribution and, thus, $F_{\left| {\varepsilon _{\rm{q}} } \right|} \left( x \right) = {\rm{erf}}\left( {{x \mathord{\left/ {\vphantom {x {\left( {\sqrt 2 \sigma _{q,{\rm{BE}}} } \right)}}} \right. \kern-\nulldelimiterspace} {\left( {\sqrt 2 \sigma _{q,{\rm{BE}}} } \right)}}} \right)$ and $1 - F_{\left| {\varepsilon _{\rm{q}} } \right|} \left( x \right) = {\rm{erfc}}\left( {{x \mathord{\left/ {\vphantom {x {\left( {\sqrt 2 \sigma _{q,{\rm{BE}}} } \right)}}} \right. \kern-\nulldelimiterspace} {\left( {\sqrt 2 \sigma _{q,{\rm{BE}}} } \right)}}} \right)$, where ${\rm{erf}}\left(  \cdot  \right)$ and ${\rm{erfc}}\left(  \cdot  \right)$ denote the error function and the complementary error function, respectively.
\subsection{Link State Modeling} \label{LinkStateModeling}
Let an arbitrary link of length $r$, \textit{i.e.}, the distance from a generic BS to the MT is equal to $r$. Motivated by recent experimental findings on mmWave channel modeling \cite[Sec. III-D]{Rappaport__mmWaveJSAC}, a three-state statistical model for each link is assumed, according to which a link can be in a LOS, NLOS or in an outage (OUT) state. A LOS state occurs if there is no blockage between BS and MT. A NLOS state, on the other hand, occurs if the BS-to-MT link is blocked. An outage state occurs if the path-loss between BS and MT is so high that no link between them can be established. In this latter case, the path-loss of the link is assumed to be infinite. In practice, outages occur implicitly when the path-loss in either a LOS or a NLOS state is sufficiently large. In \cite[Fig. 7]{Rappaport__mmWaveJSAC}, with the aid of experimental results, it is proved that adding an outage state, which is usually not observed for transmission at $\mu$Wave frequencies, provides a more accurate statistical description of the inherent coverage possibilities at mmWave frequencies.

From \cite[Eq. 8]{Rappaport__mmWaveJSAC}, the probabilities of occurrence $p_{{\rm{LOS}}} \left(  \cdot  \right)$, $p_{{\rm{NLOS}}} \left(  \cdot  \right)$, $p_{{\rm{OUT}}} \left(  \cdot  \right)$ of LOS, NLOS and outage states, respectively, as a function of the distance $r$ can be formulated as follows:
\begin{equation}
\label{Eq_4}
\begin{array}{l}
 p_{{\rm{OUT}}} \left( r \right) = \max \left\{ {0,1 - \gamma _{{\rm{OUT}}} e^{ - \delta _{{\rm{OUT}}} r} } \right\} \\
 p_{{\rm{LOS}}} \left( r \right) = \left( {1 - p_{{\rm{OUT}}} \left( r \right)} \right)\gamma _{{\rm{LOS}}} e^{ - \delta _{{\rm{LOS}}} r}  \\
 p_{{\rm{NLOS}}} \left( r \right) = \left( {1 - p_{{\rm{OUT}}} \left( r \right)} \right)\left( {1 - \gamma _{{\rm{LOS}}} e^{ - \delta _{{\rm{LOS}}} r} } \right)
 \end{array}
\end{equation}
\noindent where $\left( {\delta _{{\rm{LOS}}} ,\gamma _{{\rm{LOS}}} } \right)$ and $\left( {\delta _{{\rm{OUT}}} ,\gamma _{{\rm{OUT}}} } \right)$ are parameters that depend on the propagation scenario and on the carrier frequency being considered. Examples are available in \cite[Table I]{Rappaport__mmWaveJSAC}.

Under the assumption that the BSs are modeled as points of a homogeneous PPP and that the events that the BS-to-MT links are in LOS, NLOS or outage state are independent, $\Psi$ can be partitioned into three (one for each link state) independent and non-homogeneous PPPs, \textit{i.e.}, $\Psi _{{\rm{LOS}}}$, $\Psi _{{\rm{NLOS}}}$ and $\Psi _{{\rm{OUT}}}$, such that $\Psi  = \Psi _{{\rm{LOS}}}  \cup \Psi _{{\rm{NLOS}}}  \cup \Psi _{{\rm{OUT}}}$. This originates from the thinning property of the PPPs \cite{BaccelliBook2009}. From \eqref{Eq_4}, the densities of the PPPs $\Psi _{{\rm{LOS}}}$, $\Psi _{{\rm{NLOS}}}$ and $\Psi _{{\rm{OUT}}}$ are equal to $\lambda _{{\rm{LOS}}} \left( r \right) = \lambda p_{{\rm{LOS}}} \left( r \right)$, $\lambda _{{\rm{NLOS}}} \left( r \right) = \lambda p_{{\rm{NLOS}}} \left( r \right)$ and $\lambda _{{\rm{OUT}}} \left( r \right) = \lambda p_{{\rm{OUT}}} \left( r \right)$, respectively.
\subsection{Path-Loss Modeling} \label{PathLossModeling}
Based on the channel measurements in \cite{Rappaport__mmWaveJSAC}, the path-loss of LOS and NLOS links is as follows:
\begin{equation}
\label{Eq_5}
l_{{\rm{LOS}}} \left( r \right) = \left( {\kappa _{{\rm{LOS}}} r} \right)^{\beta _{{\rm{LOS}}} }, \quad l_{{\rm{NLOS}}} \left( r \right) = \left( {\kappa _{{\rm{NLOS}}} r} \right)^{\beta _{{\rm{NLOS}}} }
\end{equation}
\noindent where $r$ denotes a generic BS-to-MT distance, $\kappa _{{\rm{LOS}}}$ and $\kappa _{{\rm{NLOS}}}$ can be interpreted as the path-loss of LOS and NLOS links at a distance of 1 meter, respectively, $\beta _{{\rm{LOS}}}$ and $\beta _{{\rm{NLOS}}}$ denote the power path-loss exponents of LOS and NLOS links, respectively. As mentioned in Section \ref{LinkStateModeling}, the path-loss of the links that are in an outage state is assumed to be infinite, \textit{i.e.}, $l_{{\rm{OUT}}} \left( r \right) = \infty$. This model is usually known as the ``close-in'' path-loss model \cite{TED__GLOBE2013}, \cite{TED__PIMRC2014}.

The path-loss model in \eqref{Eq_5} is general enough for modeling several practical propagation conditions. For example, it can be linked to the widespread used $\left( {\alpha ,\beta } \right)$ or ``floating-intercept'' path-loss model \cite{Rappaport__mmWaveAccess}, \cite{Rappaport__mmWaveJSAC}, by setting $\kappa _{{\rm{LOS}}}  = 10^{{{\alpha _{{\rm{LOS}}} } \mathord{\left/ {\vphantom {{\alpha _{{\rm{LOS}}} } {\left( {10\beta _{{\rm{LOS}}} } \right)}}} \right. \kern-\nulldelimiterspace} {\left( {10\beta _{{\rm{LOS}}} } \right)}}}$ and $\kappa _{{\rm{NLOS}}}  = 10^{{{\alpha _{{\rm{NLOS}}} } \mathord{\left/ {\vphantom {{\alpha _{{\rm{NLOS}}} } {\left( {10\beta _{{\rm{NLOS}}} } \right)}}} \right. \kern-\nulldelimiterspace} {\left( {10\beta _{{\rm{NLOS}}} } \right)}}}$, where $\alpha _{{\rm{LOS}}}$ and $\alpha _{{\rm{NLOS}}}$ are defined in \cite[Table I]{Rappaport__mmWaveJSAC}. It is worth mentioning that in the floating-intercept model, unlike the close-in model, the parameters $\left( {\alpha ,\beta } \right)$ have no physical interpretation and they denote only the floating intercept and the slope of the best linear fit of empirical data \cite{TED__GLOBE2013}, \cite{TED__PIMRC2014}.
\subsection{Fading Modeling} \label{FadingModeling}
In addition to the distance-dependent path-loss model of Section \ref{PathLossModeling}, each link is subject to a random complex channel gain, which, for a generic BS-to-MT link, is denoted by $h$. According to \cite{Rappaport__mmWaveJSAC}, the power gain $\left| h \right|^2$ is assumed to follow a Log-Normal distribution with mean (in dB) equal to $\mu^{({\rm{dB}})}$ and standard deviation (in dB) equal to $\sigma^{({\rm{dB}})}$. Thus, $\left| h \right|^2$ takes into account large-scale shadowing. In general, $\mu^{({\rm{dB}})}$ and $\sigma^{({\rm{dB}})}$ for LOS and NLOS links are different \cite{Rappaport__mmWaveJSAC}. In what follows, they are denoted by $\mu _s^{\left( {{\rm{dB}}} \right)}$ and $\sigma _s^{\left( {{\rm{dB}}} \right)}$, where $s = \left\{ {{\rm{LOS}},{\rm{NLOS}}} \right\}$ denotes the link state.

As mentioned in Section \ref{LinkStateModeling}, for mathematical tractability, (shadowing) correlations between links are ignored. Thus, the fading power gains of LOS and NLOS links are assumed to be independent but non-identically distributed. As recently remarked and verified with the aid of simulations in \cite{Heath__mmWave}, this assumption usually causes a minor loss of accuracy in the evaluation of the statistics of the Signal-to-Interference-plus-Noise-Ratio (SINR). For ease of description, fast-fading is neglected in the present paper, but it may be readily incorporated.
\subsection{Cell Association Criterion} \label{CellAssociation}
Two cell association criteria are considered. In Section \ref{CoverageRate_PathLoss}, the MT is assumed to be served by the BS providing the smallest path-loss to it. In Section \ref{CoverageRate_Best}, the MT is assumed to be served by the BS providing the highest received power to it. In the first case study, thus, shadowing is not taken into account for cell association. The second case study, on the other hand, provides the best achievable performance at the cost of estimating large-scale shadowing \cite{Baccelli_BestAssociation}.
\subsubsection{Cell Association Based on the Smallest Path-Loss} \label{CellAssociationPathLOss}
Let $L_{{\rm{LOS}}}^{\left( 0 \right)}$, $L_{{\rm{NLOS}}}^{\left( 0 \right)}$ and $L_{{\rm{OUT}}}^{\left( 0 \right)}$ be the smallest path-loss of LOS, NLOS and OUT links, respectively. They can be formulated as:
\begin{equation}
\label{Eq_6}
L_s^{\left( 0 \right)}  = \begin{cases}
 \mathop {\min }\limits_{n \in \Psi _s } \left\{ {l_s \left( {r^{\left( n \right)} } \right)} \right\}\quad & {\rm{if}}\quad \Psi _s  \ne \emptyset  \\
  + \infty \quad & {\rm{if}}\quad \Psi _s  = \emptyset  \\
 \end{cases}, \quad L_{{\rm{OUT}}}^{\left( 0 \right)}  = \mathop {\min }\limits_{n \in \Psi _{{\rm{OUT}}} } \left\{ {l_{{\rm{OUT}}} \left( {r^{\left( n \right)} } \right)} \right\} =  + \infty
\end{equation}
\noindent where $s = \left\{ {{\mathop{\rm LOS}\nolimits} ,{\rm{NLOS}}} \right\}$, $r^{(n)}$ denotes the distance from a generic BS to the MT, and $\emptyset$ denotes an empty set. Hence, the path-loss of the serving BS, ${\rm{BS}}^{\left( 0 \right)}$, can be formulated as $L^{\left( 0 \right)}  = \min \left\{ {L_{{\rm{LOS}}}^{\left( 0 \right)} ,L_{{\rm{NLOS}}}^{\left( 0 \right)} ,L_{{\rm{OUT}}}^{\left( 0 \right)} } \right\}$.
\subsubsection{Cell Association Based on the Highest Received Power} \label{CellAssociationReceivedPower}
Let $P_{{\rm{LOS}}}^{\left( 0 \right)}$, $P_{{\rm{NLOS}}}^{\left( 0 \right)}$ and $P_{{\rm{OUT}}}^{\left( 0 \right)}$ be the inverse of the highest normalized received power of LOS, NLOS and OUT links, respectively. The received powers are normalized with respect to the transmit power of the BSs, since it is the same for all active BSs, and with respect to the directivity gain of BSs and MT, since they are the same in the absence of beamsteering errors. The impact of beamsteering errors is not taken into account during cell association. In Section \ref{MultiTierCellular}, different transmit powers and directivity gains are considered for each tier of BSs (heterogeneous cellular network) and, thus, they will be included in the cell association. Thus, $P_{{\rm{LOS}}}^{\left( 0 \right)}$, $P_{{\rm{NLOS}}}^{\left( 0 \right)}$ and $P_{{\rm{OUT}}}^{\left( 0 \right)}$ can be formulated as follows:
\begin{equation}
\label{Eq_7}
P_s^{\left( 0 \right)}  = \begin{cases}
 \mathop {\min }\limits_{n \in \Psi _s } \left\{ {{{l_s \left( {r^{\left( n \right)} } \right)} \mathord{\left/
 {\vphantom {{l_s \left( {r^{\left( n \right)} } \right)} {\left| {h_s^{\left( n \right)} } \right|^2 }}} \right.
 \kern-\nulldelimiterspace} {\left| {h_s^{\left( n \right)} } \right|^2 }}} \right\}\quad & {\rm{if}}\quad \Psi _s  \ne \emptyset  \\
  + \infty \quad & {\rm{if}}\quad \Psi _s  = \emptyset
 \end{cases}, \, P_{{\rm{OUT}}}^{\left( 0 \right)} \mathop  = \limits^{\left( a \right)} \mathop {\min }\limits_{n \in \Psi _{{\rm{OUT}}} } \left\{ {\frac{{l_{{\rm{OUT}}} \left( {r^{\left( n \right)} } \right)}}{{\left| {h_{{\rm{OUT}}}^{\left( n \right)} } \right|^2 }}} \right\} =  + \infty
\end{equation}
\noindent where a notation similar to \eqref{Eq_6} is used, $\left| {h_{\bar s}^{\left( n \right)} } \right|^2$ for $\bar s = \left\{ {{\mathop{\rm LOS}\nolimits} ,{\rm{NLOS}},{\rm{OUT}}} \right\}$ denotes the fading power gain related to LOS, NLOS and OUT BSs, respectively, and (a) holds because $\left| {h_{{\rm{OUT}}}^{\left( n \right)} } \right|^2  \ne 0$ almost surely. Hence, the inverse of the normalized received power of the serving BS, ${\rm{BS}}^{\left( 0 \right)}$, can be formulated as $P^{\left( 0 \right)}  = \min \left\{ {P_{{\rm{LOS}}}^{\left( 0 \right)} ,P_{{\rm{NLOS}}}^{\left( 0 \right)} ,P_{{\rm{OUT}}}^{\left( 0 \right)} } \right\}$.
\begin{remark} \label{Remark_Blockage}
Based on the link state model of Section \ref{LinkStateModeling}, a link may be in an outage state. Accordingly, the event that all the available BSs are in an outage state may occur with a non-zero probability. By using the notation in \eqref{Eq_6} and \eqref{Eq_7}, this occurs if $\Psi _{{\rm{LOS}}}  = \Psi _{{\rm{NLOS}}}  = \emptyset$. In this case, no BSs are available to serve the MT and it is said to be in a \textit{communication blockage state}. \hfill $\Box$
\end{remark}
\subsection{Problem Formulation} \label{ProblemFormulation}
Let $U^{\left( 0 \right)}$ be the intended received power, \textit{i.e.}, the power received at the MT and transmitted by the serving BS, ${\rm{BS}}^{\left( 0 \right)}$. If the MT is in a communication blockage state, then $U^{\left( 0 \right)} = 0$. Otherwise, $U^{\left( 0 \right)} > 0$ and it depends on the cell association being used. Thus, it is further detailed in Section \ref{CoverageRate}. The SINR of the downlink cellular network under analysis can be formulated as ${\rm{SINR}} = U^{\left( 0 \right)} \left( {\sigma _N^2  + I_{{\rm{agg}}} } \right)^{ - 1}$, where $\sigma _N^2$ is the noise power and $I_{{\rm{agg}}}$ is the aggregate other-cell interference, \textit{i.e.}, the total interference generated by the BSs in $\Psi ^{\left(\backslash 0 \right)}$. In particular, $\sigma _N^2$ is defined as $\sigma _N^2  = 10^{{{\sigma _N^2 \left( {{\rm{dBm}}} \right)} \mathord{\left/ {\vphantom {{\sigma _N^2 \left( {{\rm{dBm}}} \right)} {10}}} \right. \kern-\nulldelimiterspace} {10}}}$, where $\sigma _N^2 \left( {{\rm{dBm}}} \right) =  - 174 + 10\log _{10} \left( {{\rm{BW}}} \right) + \mathcal{F}_{{\rm{dB}}}$, ${{\rm{BW}}}$ is the transmission bandwidth and $\mathcal{F}_{{\rm{dB}}}$ is the noise figure in dB. The aggregate other-cell interference is defined as $I_{{\rm{agg}}}  = \sum\nolimits_{i \in \Psi ^{\left( {\backslash 0} \right)} } {\left( {{{\mathsf{P} G^{\left( i \right)} \left| {h^{\left( i \right)} } \right|^2 } \mathord{\left/ {\vphantom {{\mathsf{P} G^{\left( i \right)} \left| {h^{\left( i \right)} } \right|^2 } {L^{\left( i \right)} }}} \right. \kern-\nulldelimiterspace} {  l\left( {r^{\left( i \right)} } \right)    }}} \right)}$, where $\mathsf{P}$ is the transmit power of the BSs and $l\left(  \cdot  \right)$ is the path-loss of Section \ref{PathLossModeling}, which depends on a BS being in a LOS, NLOS or outage state.

From the SINR, coverage probability (${\rm{P}}_{{\mathop{\rm cov}} }$) and average rate ($\rm{R}$) can be formulated as \cite{MDR_COMMLPeng}:
\begin{equation}
\label{Eq_8}
{\rm{P}}^{({\mathop{\rm cov}} )} \left( {\rm{T}} \right) = \Pr \left\{ {{\rm{SINR}} \ge {\rm{T}}} \right\}
\end{equation} 
\begin{equation}
\label{Eq_9}
\begin{split}
 {\rm{R}} &= {\mathbb{E}}_{{\rm{SINR}}} \left\{ {{\rm{BW}}\log _2 \left( {1 + {\rm{SINR}}} \right)} \right\}
  = \frac{{{\rm{BW}}}}{{\ln \left( 2 \right)}}\int\nolimits_0^{ + \infty } {{\rm{P}}^{\left( {{\mathop{\rm cov}} } \right)} \left( {e^t  - 1} \right)dt} \\ &  = \frac{{{\rm{BW}}}}{{\ln \left( 2 \right)}}\int\nolimits_0^{ + \infty } {\frac{{{\rm{P}}^{\left( {{\mathop{\rm cov}} } \right)} \left( t \right)}}{{t + 1}}dt} \mathop  \approx \limits^{\left( a \right)} \frac{{{\rm{BW}}}}{{\ln \left( 2 \right)}}\sum\limits_{u = 1}^{{\rm{N}}_{{\rm{GCQ}}} } {w^{\left( u \right)} \frac{{{\rm{P}}^{\left( {{\mathop{\rm cov}} } \right)} \left( {x^{\left( u \right)} } \right)}}{{x^{\left( u \right)}  + 1}}}  \\
 \end{split}
\end{equation}
\noindent where ${\rm{T}} > 0$ is a reliability threshold and ${\mathbb{E}}\left\{  \cdot  \right\}$ denotes the expectation operator. The approximation in (a) follows from the Gauss-Chebyshev Quadrature (GCQ) rule \cite[Eq. (25.4.39)]{AbramowitzStegun}, where ${w^{\left( u \right)} }$ and ${x^{\left( u \right)} }$ for $u=1,2,\ldots, {{\rm{N}}_{{\rm{GCQ}}} }$ are weights and abscissas of the quadrature, respectively, which are available in closed-form in \cite[Eq. (13)]{MDR_TCOMrate}. The approximation in (a) is especially useful when the coverage probability cannot be formulated in a closed-form expression.

In the next sections, new mathematical expressions for ${\rm{P}}^{({\mathop{\rm cov}})}$ are provided. The analytical formulation is based on the noise-limited approximation of mmWave cellular communications, \textit{i.e.}, ${\rm{SINR}} \approx {\rm{SNR}} = {{U^{\left( 0 \right)} } \mathord{\left/ {\vphantom {{U^{\left( 0 \right)} } {\sigma _N^2 }}} \right. \kern-\nulldelimiterspace} {\sigma _N^2 }}$, which has been observed in recent studies, both with the aid of numerical simulations and field measurements \cite{Rappaport__mmWaveJSAC}, \cite{Singh__mmWave}, \cite{PIMRC_2014}. In Section \ref{Results}, the validity and the accuracy of the noise-limited approximation are substantiated with the aid of Monte Carlo simulations, which account for the other-cell interference as well. Therefore, for simplicity, in the rest of the manuscript the SINR is not used anymore and only the SNR is considered. From the coverage probability, the average rate is obtained from \eqref{Eq_9}. Thus, less emphasis is put on it.
\begin{remark} \label{Remark_BlockageCoverageRate}
If a communication blockage occurs, \textit{i.e.}, $U^{\left( 0 \right)} = 0$, then ${\rm{SNR}} = 0$ and coverage probability and average rate are equal to zero. The coverage is zero regardless of ${\rm{T}} > 0$. \hfill $\Box$
\end{remark}
\section{Analysis and Approximations of Transformations of the Path-Loss} \label{PPP_PathLoss}
In this section, we provide general results for the distribution of transformations of the path-loss of mmWave systems, which account for LOS, NLOS and outage states. These results are useful for computing the coverage probability and the average rate in Section \ref{CoverageRate}.
\begin{lemma} \label{Intensity_Lemma}
Let $\Phi  = \left\{ {\Phi _{{\rm{LOS}}} ,\Phi _{{\rm{NLOS}}} ,\Phi _{{\rm{OUT}}} } \right\}$, where $\Phi _{{\rm{LOS}}}  = \left\{ {{{l_{{\rm{LOS}}} \left( {r^{\left( n \right)} } \right)} \mathord{\left/ {\vphantom {{l_{{\rm{LOS}}} \left( {r^{\left( n \right)} } \right)} {{\mathcal{A}}_{{\rm{LOS}}}^{\left( n \right)} }}} \right. \kern-\nulldelimiterspace} {{\mathcal{A}}_{{\rm{LOS}}}^{\left( n \right)} }},n \in \Psi _{{\rm{LOS}}} } \right\}$, $\Phi _{{\rm{NLOS}}}  = \left\{ {{{l_{{\rm{NLOS}}} \left( {r^{\left( n \right)} } \right)} \mathord{\left/ {\vphantom {{l_{{\rm{NLOS}}} \left( {r^{\left( n \right)} } \right)} {{\mathcal{A}}_{{\rm{NLOS}}}^{\left( n \right)} }}} \right. \kern-\nulldelimiterspace} {{\mathcal{A}}_{{\rm{NLOS}}}^{\left( n \right)} }},n \in \Psi _{{\rm{NLOS}}} } \right\}$ and $\Phi _{{\rm{OUT}}}  = \left\{ {{{l_{{\rm{OUT}}} \left( {r^{\left( n \right)} } \right)} \mathord{\left/ {\vphantom {{l_{{\rm{OUT}}} \left( {r^{\left( n \right)} } \right)} {{\mathcal{A}}_{{\rm{OUT}}}^{\left( n \right)} }}} \right. \kern-\nulldelimiterspace} {{\mathcal{A}}_{{\rm{OUT}}}^{\left( n \right)} }},n \in \Psi _{{\rm{OUT}}} } \right\}$ are transformations of the path-loss of LOS, NLOS and OUT BSs, respectively, which is defined in Sections \ref{LinkStateModeling} and \ref{PathLossModeling}. Let ${\mathcal{A}}_{\bar s}^{\left( n \right)}$ for $n \in \Psi _{\bar s}$ and $\bar s \in \left\{ {{\rm{LOS}},{\rm{NLOS}},{\rm{OUT}}} \right\}$ be:
\begin{enumerate}
\item A set of equal constants, \textit{i.e.}, ${\mathcal{A}}_{\bar s}^{\left( n \right)} = {\mathcal{A}}_{\bar s}$ for $n \in \Psi _{\bar s}$, or
\item a set of independent and identically distributed random variables with ${\mathcal{A}}_{\bar s}$ denoting a random variable having the same distribution as any ${\mathcal{A}}_{\bar s}^{\left( n \right)}$ for $n \in \Psi _{\bar s}$.
\end{enumerate}

Let the link state model in \eqref{Eq_4}. Then, $\Phi$ is a PPP with intensity as follows:
\begin{equation}
\label{Eq_11}
\Lambda _\Phi  \left( {\left[ {0,x} \right)} \right) = \begin{cases}
   \tilde \Lambda _\Phi  \left( {\left[ {0,x} \right)} \right)\quad & {\rm{if}}\quad \left( {{\mathcal{A}}_{{\rm{LOS}}} ,{\mathcal{A}}_{{\rm{NLOS}}} } \right)\;{\rm{are \, constants}} \hfill  \\
   \bar \Lambda _\Phi  \left( {\left[ {0,x} \right)} \right)\quad & {\rm{if}}\quad \left( {{\mathcal{A}}_{{\rm{LOS}}} ,{\mathcal{A}}_{{\rm{NLOS}}} } \right)\;{\rm{are \, random \, variables}} \hfill  \\
\end{cases}
\end{equation}
\noindent where the following definitions hold:
\begin{equation}
\label{Eq_11bis}
\begin{split}
 & \hspace{-0.40cm} \tilde \Lambda _\Phi  \left( {\left[ {0,x} \right)} \right) = \Lambda _{{\rm{LOS}}} \left( {\left[ {0,{\mathcal{A}}_{{\rm{LOS}}} x} \right)} \right) + \Lambda _{{\rm{NLOS}}} \left( {\left[ {0,{\mathcal{A}}_{{\rm{NLOS}}} x} \right)} \right) \\
 & \hspace{-0.40cm} \bar \Lambda _\Phi  \left( {\left[ {0,x} \right)} \right) = {\mathbb{E}}_{{\mathcal{A}}_{{\rm{LOS}}} } \left\{ {\Lambda _{{\rm{LOS}}} \left( {\left[ {0,{\mathcal{A}}_{{\rm{LOS}}} x} \right)} \right)} \right\} + {\mathbb{E}}_{{\mathcal{A}}_{{\rm{NLOS}}} } \left\{ {\Lambda _{{\rm{NLOS}}} \left( {\left[ {0,{\mathcal{A}}_{{\rm{NLOS}}} x} \right)} \right)} \right\} \\
 & \hspace{-0.40cm} \Lambda _{{\rm{LOS}}} \left( {\left[ {0,x} \right)} \right) = \Upsilon _0 \left( {x;s = {\rm{LOS}}} \right),\; \Lambda _{{\rm{NLOS}}} \left( {\left[ {0,x} \right)} \right) = \Upsilon _1 \left( {x;s = {\rm{NLOS}}} \right) - \Upsilon _0 \left( {x;s = {\rm{NLOS}}} \right) \\
 \end{split}
\end{equation}
\begin{equation} 
\label{Eq_12}
\begin{split}
 & \hspace{-0.25cm} \Upsilon _0 \left( x;s \right) = {\mathcal{K}}_2 \left( {e^{ - W }  + W e^{ - W }  - e^{ - V_s x^{{1 \mathord{\left/
 {\vphantom {1 {\beta _s }}} \right.
 \kern-\nulldelimiterspace} {\beta _s }}} }  - V_s x^{{1 \mathord{\left/
 {\vphantom {1 {\beta _s }}} \right.
 \kern-\nulldelimiterspace} {\beta _s }}} e^{ - V_s x^{{1 \mathord{\left/
 {\vphantom {1 {\beta _s }}} \right.
 \kern-\nulldelimiterspace} {\beta _s }}} } } \right){\mathcal{H}}\left( {x - Z_s } \right) \\
 & \hspace{-0.25cm} + {\mathcal{K}}_1 \left( {1 - e^{ - Q_s x^{{1 \mathord{\left/
 {\vphantom {1 {\beta _s }}} \right.
 \kern-\nulldelimiterspace} {\beta _s }}} }  - Q_s x^{{1 \mathord{\left/
 {\vphantom {1 {\beta _s }}} \right.
 \kern-\nulldelimiterspace} {\beta _s }}} e^{ - Q_s x^{{1 \mathord{\left/
 {\vphantom {1 {\beta _s }}} \right.
 \kern-\nulldelimiterspace} {\beta _s }}} } } \right){\mathcal{\bar H}}\left( {x - Z_s } \right)
   + {\mathcal{K}}_1 \left( {1 - e^{ - R }  - R e^{ - R } } \right){\mathcal{H}}\left( {x - Z_s } \right) \\
 & \hspace{-0.25cm} \Upsilon _1 \left( {x;s} \right) = \pi \lambda \kappa _s^{ - 2} x^{{2 \mathord{\left/
 {\vphantom {2 {\beta _s }}} \right.
 \kern-\nulldelimiterspace} {\beta _s }}} {\mathcal{\bar H}}\left( {x - Z_s } \right) + \pi \lambda \left( {\delta _{{\rm{OUT}}}^{ - 1} \ln \left( {\gamma _{{\rm{OUT}}} } \right)} \right)^2 {\mathcal{H}}\left( {x - Z_s } \right) \\
  & \hspace{-0.25cm} + 2\pi \lambda \delta _{{\rm{OUT}}}^{ - 2} \gamma _{{\rm{OUT}}} \left( {\gamma _{{\rm{OUT}}}^{ - 1}  + \gamma _{{\rm{OUT}}}^{ - 1} \ln \left( {\gamma _{{\rm{OUT}}} } \right) - e^{ - T_s x^{{1 \mathord{\left/
 {\vphantom {1 {\beta _s }}} \right.
 \kern-\nulldelimiterspace} {\beta _s }}} }  - T_s x^{{1 \mathord{\left/
 {\vphantom {1 {\beta _s }}} \right.
 \kern-\nulldelimiterspace} {\beta _s }}} e^{ - T_s x^{{1 \mathord{\left/
 {\vphantom {1 {\beta _s }}} \right.
 \kern-\nulldelimiterspace} {\beta _s }}} } } \right){\mathcal{H}}\left( {x - Z_s } \right) \\
 \end{split}
\end{equation}
\noindent where ${\mathcal{H}}\left(  \cdot  \right)$ is the Heaviside function, ${\mathcal{\bar H}}\left( x \right) = 1 - {\mathcal{H}}\left( x \right)$, ${\mathcal{K}}_1  = 2\pi \lambda \gamma _{{\rm{LOS}}} \delta _{{\rm{LOS}}}^{ - 2}$, ${\mathcal{K}}_2  = 2\pi \lambda \gamma _{{\rm{LOS}}} \gamma _{{\rm{OUT}}}$ $\times \left( {\delta _{{\rm{LOS}}}  + \delta _{{\rm{OUT}}} } \right)^{ - 2}$, $R  = \delta _{{\rm{LOS}}} \delta _{{\rm{OUT}}}^{ - 1} \ln \left( {\gamma _{{\rm{OUT}}} } \right)$, $W  = \left( {\delta _{{\rm{LOS}}}  + \delta _{{\rm{OUT}}} } \right)\delta _{{\rm{OUT}}}^{ - 1} \ln \left( {\gamma _{{\rm{OUT}}} } \right)$, $Q_s  = \delta _{{\rm{LOS}}} \kappa _s^{ - 1}$, $T_s  = \delta _{{\rm{OUT}}} \kappa _s^{ - 1}$, $V_s  = \left( {\delta _{{\rm{LOS}}}  + \delta _{{\rm{OUT}}} } \right)\kappa _s^{ - 1}$, $Z_s  = \left( {\kappa _s \delta _{{\rm{OUT}}}^{ - 1} \ln \left( {\gamma _{{\rm{OUT}}} } \right)} \right)^{\beta _s }$ for $s = \left\{ {{\rm{LOS}},{\rm{NLOS}}} \right\}$.

\emph{Proof}: See Appendix I. \hfill $\Box$
\end{lemma}
\begin{corollary} \label{CDF_Corollary}
Let $\delta _{{\rm{OUT}}}  = 0$, $\gamma _{{\rm{OUT}}}  = 1$, \textit{i.e.}, $p_{{\rm{OUT}}} \left( r \right) = 0$ in \eqref{Eq_4}. $\Lambda _\Phi  \left(  \cdot  \right)$ in \eqref{Eq_11} holds with $\Upsilon _0 \left( x;s \right) = {\mathcal{K}}_1 \left( {1 - e^{ - Q_s x^{{1 \mathord{\left/ {\vphantom {1 {\beta _s }}} \right. \kern-\nulldelimiterspace} {\beta _s }}} }  - Q_s x^{{1 \mathord{\left/ {\vphantom {1 {\beta _s }}} \right. \kern-\nulldelimiterspace} {\beta _s }}} e^{ - Q_s x^{{1 \mathord{\left/ {\vphantom {1 {\beta _s }}} \right. \kern-\nulldelimiterspace} {\beta _s }}} } } \right)$, $\Upsilon _1 \left( {x;s} \right) = \pi \lambda \kappa _s^{ - 2} x^{{2 \mathord{\left/ {\vphantom {2 {\beta _s }}} \right. \kern-\nulldelimiterspace} {\beta _s }}}$, $s = \left\{ {{\rm{LOS}},{\rm{NLOS}}} \right\}$.

\emph{Proof}: It follows directly from \eqref{Eq_12}, since $Z_s \to +\infty$ for $s = \left\{ {{\rm{LOS}},{\rm{NLOS}}} \right\}$. \hfill $\Box$
\end{corollary}
\begin{lemma} \label{CDF_Lemma}
Let $\Phi ^{\left( 0 \right)}  = \min \left\{ \Phi  \right\}$ be the smallest element of the PPP $\Phi$ introduced in \textit{Lemma \ref{Intensity_Lemma}}. Its CDF, \textit{i.e.}, $F_{\Phi^{\left( 0 \right)} } \left( x \right) = \Pr \left\{ {\Phi^{\left( 0 \right)}  \le x} \right\}$, can be formulated as follows:
\begin{equation}
\label{Eq_11cdf}
F_{\Phi^{\left( 0 \right)} } \left( x \right) = 1 - \exp \left( { - \Lambda _\Phi \left( {\left[ {0,x} \right)} \right)} \right)
\end{equation}
\noindent where $\Lambda _\Phi  \left( \cdot \right)$ is defined in \eqref{Eq_11}.

\emph{Proof}: It follows by applying the void probability theorem of PPPs \cite[Corollary 6]{Blaszczyszyn_Infocom2013}. \hfill $\Box$
\end{lemma}
\begin{remark} \label{Remark_GeneralPPP}
The transformation of the path-loss in \textit{Lemma \ref{Intensity_Lemma}} accounts for cell associations based on both the smallest path-loss and the highest received power. In particular, the first case study is obtained by setting ${\left( {{\mathcal{A}}_{{\rm{LOS}}} ,{\mathcal{A}}_{{\rm{NLOS}}} } \right)} = {\left( 1,1 \right)}$, while the second case study follows by setting $\left( {{\mathcal{A}}_{{\rm{LOS}}}^{\left( n \right)} ,{\mathcal{A}}_{{\rm{NLOS}}}^{\left( n \right)} } \right) = \left( {\left| {h_{{\rm{LOS}}}^{\left( n \right)} } \right|^2 ,\left| {h_{{\rm{NLOS}}}^{\left( n \right)} } \right|^2 } \right)$ for $n \in \Psi _{{\rm{LOS}}}$ and $n \in \Psi _{{\rm{NLOS}}}$. \hfill $\Box$
\end{remark}
\begin{remark} \label{Remark_AgainstBartek}
In \cite{Blaszczyszyn_Infocom2013}, the intensity of the PPP of the path-loss is computed under the assumption of a single-state link model, \textit{i.e.}, no outage state exists and the distributions of LOS and NLOS links are the same. Thus, \textit{Lemma \ref{Intensity_Lemma}} generalizes the results in \cite{Blaszczyszyn_Infocom2013}, by taking into account the peculiarities of mmWave communications. Also, it reduces to \cite{Blaszczyszyn_Infocom2013} under the same assumptions. In \cite{Blaszczyszyn_Infocom2013}, it is shown that the impact of Log-Normal shadowing on the cell association based on the highest received power criterion consists of a scaling factor of the PPP density, which is a function of the fractional moments of the Log-Normal distribution. \textit{Lemma \ref{Intensity_Lemma}}, on the other hand, shows that Log-Normal shadowing has a more complicated impact in mmWave systems. \hfill $\Box$
\end{remark}
\subsection{Two-Ball Approximation} \label{TwoBall_Approximation}
From \eqref{Eq_11} and \eqref{Eq_11bis}, it is apparent that the intensity of $\Phi$ is available in closed-form if ${\mathcal{A}}_{{\rm{LOS}}}$ and ${\mathcal{A}}_{{\rm{NLOS}}}$ are constants. The expectation in \eqref{Eq_11bis}, on the other hand, needs to be computed if they are random variables. To the best of the author's knowledge, however, the expectation in \eqref{Eq_11bis} cannot be computed in closed-form if the channel power gains follow a Log-Normal distribution. The computation of the expectation may be possible, however, for other fading distributions. This originates from the mathematical intractability of the Log-Normal distribution and from the fact that no closed-form expression for its Laplace transform exists. In order to overcome this issue, we propose an approximation for modeling the state of links in LOS, NLOS and outage.

The proposed approach consists of computing the link state probabilities based on a ``two-ball'' approximation of \eqref{Eq_4}. More specifically, the probabilities in \eqref{Eq_4} are approximated as follows:
\begin{equation}
\label{Eq_2Ball__1}
\left\{ \begin{split}
 & p_{\bar s} \left( r \right) \approx  p_{\bar s}^{\left( {{\rm{approx}}} \right)} \left( r \right) = q_{\bar s}^{\left[ {0,D_1 } \right]} {\bf{1}}_{\left[ {0,D_1 } \right)} \left( r \right) + q_{\bar s}^{\left[ {D_1 ,D_2 } \right]} {\bf{1}}_{\left( {D_1 ,D_2 } \right)} \left( r \right) + q_{\bar s}^{\left[ {D_2 ,\infty } \right]} {\bf{1}}_{\left[ {D_2 , + \infty } \right)} \left( r \right) \\
 & \sum\limits_{\bar s \in \left\{ {{\rm{LOS}},{\rm{NLOS}},{\rm{OUT}}} \right\}} {q_{\bar s}^{\left[ {0,D_1 } \right]} }  = \sum\limits_{\bar s \in \left\{ {{\rm{LOS}},{\rm{NLOS}},{\rm{OUT}}} \right\}} {q_{\bar s}^{\left[ {D_1 ,D_2 } \right]} }  = \sum\limits_{\bar s \in \left\{ {{\rm{LOS}},{\rm{NLOS}},{\rm{OUT}}} \right\}} {q_{\bar s}^{\left[ {D_2 ,\infty } \right]} } = 1  \\
 \end{split} \right.
\end{equation}
\noindent where $\bar s \in \left\{ {{\rm{LOS}},{\rm{NLOS}},{\rm{OUT}}} \right\}$, $D_2  \ge D_1  \ge 0$ are the radii of the approximating balls, ${\bf{1}}_{\left[ {a,b} \right)} \left(  \cdot  \right)$ is the indicator function, which is defined as ${\bf{1}}_{\left[ {a,b} \right)} \left( r \right) = 1$ if $r \in \left[ {a,b} \right)$ and ${\bf{1}}_{\left[ {a,b} \right)} \left( r \right) = 0$ if $r \notin \left[ {a,b} \right)$, and $q_{\bar s}^{\left[ {a,b} \right]}$ denotes the probability that a link of length $r \in \left[ {a,b} \right)$ is in state $\bar s$. The second equality in \eqref{Eq_2Ball__1} guarantees that each link of length $r$ is only in one of the three possible states $\bar s \in \left\{ {{\rm{LOS}},{\rm{NLOS}},{\rm{OUT}}} \right\}$. In what follows, it is referred to as \textit{approximation constraint}.
\begin{remark} \label{Remark_2BallRationale}
The rationale behind \eqref{Eq_2Ball__1} originates from the visual inspection of \cite[Fig. 7]{Rappaport__mmWaveJSAC}. It is apparent from \cite[Fig. 7]{Rappaport__mmWaveJSAC}, in fact, that two breaking distances ($D_1$ and $D_2$) emerge for arbitrary values of the link length $r$, which result in three connectivity regions: the first, for $r \in \left[ {0,D_1 } \right)$, where the links are most likely to be either in LOS or NLOS; the second, for $r \in \left( {D_1,D_2 } \right)$, where the links can be in any state; and the third, for ${\left[ {D_2 , + \infty } \right)}$, where the links are most likely to be in outage. Equation \eqref{Eq_2Ball__1} accounts for this empirical observation for any $q_{\bar s}^{\left[ {a,b} \right]}$. \hfill $\Box$
\end{remark}
\begin{remark} \label{Remark_2BallNovelty}
The two-ball approximation in \eqref{Eq_2Ball__1} may be seen as a generalization of the single-ball approximation introduced in \cite{Heath__Blockage}, \cite{Singh__mmWave}. Compared to these papers, in particular, it accounts for the outage state that emerges in mmWave communications. In addition, the approach for estimating the parameters of the approximation is different and based on a technique introduced in the present paper for the first time, which is referred to as \textit{path-loss intensity matching}. \hfill $\Box$
\end{remark}

Before describing the path-loss intensity matching approach for computing the parameters of the approximation in \eqref{Eq_2Ball__1}, \textit{i.e.}, $\left( { D_1 ,D_2 ,q_{\bar s}^{\left[ {0,D_1 } \right]} ,q_{\bar s}^{\left[ {D_1 ,D_2 } \right]} ,q_{\bar s}^{\left[ {D_2 ,\infty } \right]} } \right)$ for $\bar s \in \left\{ {{\rm{LOS}},{\rm{NLOS}},{\rm{OUT}}} \right\}$, \textit{Lemma \ref{Intensity_Lemma}} needs to be generalized based on the link state model in \eqref{Eq_2Ball__1}.
\begin{lemma} \label{Intensity2Ball_Lemma}
Let $\Phi ^{\left( {{\rm{approx}}} \right)}  = \left\{ {\Phi _{{\rm{LOS}}}^{\left( {{\rm{approx}}} \right)} ,\Phi _{{\rm{NLOS}}}^{\left( {{\rm{approx}}} \right)} ,\Phi _{{\rm{OUT}}}^{\left( {{\rm{approx}}} \right)} } \right\}$, where $\Phi _{{\rm{LOS}}}^{\left( {{\rm{approx}}} \right)}  = \left\{ {{l_{{\rm{LOS}}} \left( {r^{\left( n \right)} } \right)} \mathord{\left/ {\vphantom {{l_{{\rm{LOS}}} \left( {r^{\left( n \right)} } \right)} {{\mathcal{A}}_{{\rm{LOS}}}^{\left( n \right)} }}} \right. \kern-\nulldelimiterspace} {{\mathcal{A}}_{{\rm{LOS}}}^{\left( n \right)} }}, \right.$ $\left. n \in \Psi _{{\rm{LOS}}}^{\left( {{\rm{approx}}} \right)} \right\}$, $\Phi _{{\rm{NLOS}}}^{\left( {{\rm{approx}}} \right)}  = \left\{ {{{l_{{\rm{NLOS}}} \left( {r^{\left( n \right)} } \right)} \mathord{\left/ {\vphantom {{l_{{\rm{NLOS}}} \left( {r^{\left( n \right)} } \right)} {{\mathcal{A}}_{{\rm{NLOS}}} }}} \right. \kern-\nulldelimiterspace} {{\mathcal{A}}_{{\rm{NLOS}}} }},n \in \Psi _{{\rm{NLOS}}}^{\left( {{\rm{approx}}} \right)} } \right\}$, $\Phi _{{\rm{OUT}}}^{\left( {{\rm{approx}}} \right)}  = \left\{ {{l_{{\rm{OUT}}} \left( {r^{\left( n \right)} } \right)} \mathord{\left/ {\vphantom {{l_{{\rm{OUT}}} \left( {r^{\left( n \right)} } \right)} {{\mathcal{A}}_{{\rm{OUT}}}^{\left( n \right)} }}} \right. \kern-\nulldelimiterspace} {{\mathcal{A}}_{{\rm{OUT}}}^{\left( n \right)} }}, \right.$ $\left. n \in \Psi _{{\rm{OUT}}}^{\left( {{\rm{approx}}} \right)} \right\}$ are transformations of the path-loss of LOS, NLOS and OUT BSs, respectively, where the path-loss model is defined in Section \ref{PathLossModeling} and the link state model is given by \eqref{Eq_2Ball__1}, \textit{i.e.}, $\Psi _{\bar s}^{\left( {{\rm{approx}}} \right)}$ for $\bar s \in \left\{ {{\rm{LOS}},{\rm{NLOS}},{\rm{OUT}}} \right\}$ has the same definition as $\Psi _s$ except that \eqref{Eq_4} is replaced by \eqref{Eq_2Ball__1}. Let ${\mathcal{A}}_{\bar s}^{\left( n \right)}$ for $n \in \Psi _{\bar s}$ and $\bar s \in \left\{ {{\rm{LOS}},{\rm{NLOS}},{\rm{OUT}}} \right\}$ be defined as in \textit{Lemma \ref{Intensity_Lemma}}. Then, $\Phi ^{\left( {{\rm{approx}}} \right)}$ is a PPP with intensity given in \eqref{Eq_11} and \eqref{Eq_11bis}, which are obtained by replacing $\Lambda _s \left( { \cdot } \right)$ for $s \in \left\{ {{\rm{LOS}},{\rm{NLOS}}} \right\}$ with $\Lambda _s^{\left( {{\rm{approx}}} \right)} \left( { \cdot } \right)$ defined as follows:
\begin{equation}
\label{Eq_2Ball__2}
\begin{split}
\hspace{-0.25cm} \Lambda _s^{\left( {{\rm{approx}}} \right)} \left( {\left[ {0,x} \right)} \right)  &=  - {\mathcal{G}}_s^{\left( 3 \right)}  + \left( {{\mathcal{G}}_s^{\left( 1 \right)} x^{{2 \mathord{\left/
 {\vphantom {2 {\beta _s }}} \right.
 \kern-\nulldelimiterspace} {\beta _s }}}  + {\mathcal{G}}_s^{\left( 3 \right)} } \right){\mathcal{\bar H}}\left( {x - \left( {\kappa _s D_1 } \right)^{\beta _s } } \right) + {\mathcal{G}}_s^{\left( 4 \right)} {\mathcal{H}}\left( {x - \left( {\kappa _s D_1 } \right)^{\beta _s } } \right) \\
  &+ {\mathcal{G}}_s^{\left( 2 \right)} x^{{2 \mathord{\left/
 {\vphantom {2 {\beta _s }}} \right.
 \kern-\nulldelimiterspace} {\beta _s }}} {\mathcal{\bar H}}\left( {x - \left( {\kappa _s D_2 } \right)^{\beta _s } } \right) + \left( {{\mathcal{G}}_s^{\left( 6 \right)} x^{{2 \mathord{\left/
 {\vphantom {2 {\beta _s }}} \right.
 \kern-\nulldelimiterspace} {\beta _s }}}  + {\mathcal{G}}_s^{\left( 5 \right)} } \right){\mathcal{H}}\left( {x - \left( {\kappa _s D_2 } \right)^{\beta _s } } \right) \\
 \end{split}
\end{equation}
\noindent where ${\mathcal{G}}_s^{\left( 1 \right)}  = \pi \lambda \kappa _s^{ - 2} \left( {q_s^{\left[ {0,D_1 } \right]}  - q_s^{\left[ {D_1 ,D_2 } \right]} } \right)$, ${\mathcal{G}}_s^{\left( 2 \right)}  = \pi \lambda \kappa _s^{ - 2} q_s^{\left[ {D_1 ,D_2 } \right]}$, ${\mathcal{G}}_s^{\left( 3 \right)}  = \pi \lambda D_1^2 q_s^{\left[ {D_1 ,D_2 } \right]}$, ${\mathcal{G}}_s^{\left( 4 \right)}  = \pi \lambda D_1^2 q_s^{\left[ {0,D_1 } \right]}$, ${\mathcal{G}}_s^{\left( 5 \right)}  = \pi \lambda D_2^2 \left( {q_s^{\left[ {D_1 ,D_2 } \right]}  - q_s^{\left[ {D_2 ,\infty } \right]} } \right)$ and ${\mathcal{G}}_s^{\left( 6 \right)}  = \pi \lambda \kappa _s^{ - 2} q_s^{\left[ {D_2 ,\infty } \right]}$.

\emph{Proof}: The proof follows the same steps as the proof of \textit{Lemma \ref{Intensity_Lemma}}. The only difference lies in replacing $p_s \left(  \cdot  \right)$ in \eqref{Eq_App1__1} with $p_{s}^{\left( {{\rm{approx}}} \right)} \left( \cdot \right)$ in \eqref{Eq_2Ball__1} and by computing the related integrals. \hfill $\Box$
\end{lemma}
\begin{corollary} \label{Intensity2Ball_Corollary}
Let $\Lambda _s^{\left( {{\rm{approx}}} \right)} \left( { \cdot } \right)$ in \eqref{Eq_2Ball__2} for $s \in \left\{ {{\rm{LOS}},{\rm{NLOS}}} \right\}$. Let ${\mathcal{A}}_s$ be a Log-Normal random variable with mean (in dB) and standard deviation (in dB) equal to $\mu _s^{\left( {{\rm{dB}}} \right)}$ and $\sigma _s^{\left( {{\rm{dB}}} \right)}$, respectively. Then, the following holds for $s \in \left\{ {{\rm{LOS}},{\rm{NLOS}}} \right\}$:
\begin{equation}
\label{Eq_2Ball__3}
\hspace{-0.85cm} \begin{split}
 & \bar \Lambda _s^{\left( {{\rm{approx}}} \right)} \left( {\left[ {0,x} \right)} \right) = {\mathbb{E}}_{{\mathcal{A}}_s } \left\{ {\Lambda _s^{\left( {{\rm{approx}}} \right)} \left( {\left[ {0,{\mathcal{A}}_s x} \right)} \right)} \right\} \\ & = {\mathcal{G}}_s^{\left( 1 \right)} x^{{2 \mathord{\left/
 {\vphantom {2 {\beta _s }}} \right.
 \kern-\nulldelimiterspace} {\beta _s }}} m_{{\mathcal{A}}_s } \left( {{2 \mathord{\left/
 {\vphantom {2 {\beta _s }}} \right.
 \kern-\nulldelimiterspace} {\beta _s }},{{\left( {\kappa _s D_1 } \right)^{\beta _s } } \mathord{\left/
 {\vphantom {{\left( {\kappa _s D_1 } \right)^{\beta _s } } x}} \right.
 \kern-\nulldelimiterspace} x}} \right) + {\mathcal{G}}_s^{\left( 2 \right)} x^{{2 \mathord{\left/
 {\vphantom {2 {\beta _s }}} \right.
 \kern-\nulldelimiterspace} {\beta _s }}} m_{{\mathcal{A}}_s } \left( {{2 \mathord{\left/
 {\vphantom {2 {\beta _s }}} \right.
 \kern-\nulldelimiterspace} {\beta _s }},{{\left( {\kappa _s D_2 } \right)^{\beta _s } } \mathord{\left/
 {\vphantom {{\left( {\kappa _s D_2 } \right)^{\beta _s } } x}} \right.
 \kern-\nulldelimiterspace} x}} \right)
  - {\mathcal{G}}_s^{\left( 3 \right)} \bar F_{{\mathcal{A}}_s } \left( {{{\left( {\kappa _s D_1 } \right)^{\beta _s } } \mathord{\left/
 {\vphantom {{\left( {\kappa _s D_1 } \right)^{\beta _s } } x}} \right.
 \kern-\nulldelimiterspace} x}} \right) \\ & + {\mathcal{G}}_s^{\left( 4 \right)} \bar F_{{\mathcal{A}}_s } \left( {{{\left( {\kappa _s D_1 } \right)^{\beta _s } } \mathord{\left/
 {\vphantom {{\left( {\kappa _s D_1 } \right)^{\beta _s } } x}} \right.
 \kern-\nulldelimiterspace} x}} \right) + {\mathcal{G}}_s^{\left( 5 \right)} \bar F_{{\mathcal{A}}_s } \left( {{{\left( {\kappa _s D_2 } \right)^{\beta _s } } \mathord{\left/
 {\vphantom {{\left( {\kappa _s D_2 } \right)^{\beta _s } } x}} \right.
 \kern-\nulldelimiterspace} x}} \right) + {\mathcal{G}}_s^{\left( 6 \right)} x^{{2 \mathord{\left/
 {\vphantom {2 {\beta _s }}} \right.
 \kern-\nulldelimiterspace} {\beta _s }}} \bar m_{{\mathcal{A}}_s } \left( {{2 \mathord{\left/
 {\vphantom {2 {\beta _s }}} \right.
 \kern-\nulldelimiterspace} {\beta _s }},{{\left( {\kappa _s D_2 } \right)^{\beta _s } } \mathord{\left/
 {\vphantom {{\left( {\kappa _s D_2 } \right)^{\beta _s } } x}} \right.
 \kern-\nulldelimiterspace} x}} \right) \\
 \end{split}
\end{equation}
\noindent where $m_{{\mathcal{A}}_s } \left( {\nu ,y} \right) = \left( {{1 \mathord{\left/ {\vphantom {1 2}} \right. \kern-\nulldelimiterspace} 2}} \right)\exp \left\{ {\nu \mu _s  + \left( {{1 \mathord{\left/ {\vphantom {1 2}} \right. \kern-\nulldelimiterspace} 2}} \right)\nu ^2 \sigma _s^2 } \right\}{\rm{erfc}}\left( {{{\nu \sigma _s } \mathord{\left/ {\vphantom {{\nu \sigma _s } {\sqrt 2 }}} \right. \kern-\nulldelimiterspace} {\sqrt 2 }} - {{\left( {\ln \left( y \right) - \mu _s } \right)} \mathord{\left/ {\vphantom {{\left( {\ln \left( y \right) - \mu _s } \right)} {\left( {\sqrt 2 \sigma _s } \right)}}} \right. \kern-\nulldelimiterspace} {\left( {\sqrt 2 \sigma _s } \right)}}} \right)$, $F_{{\mathcal{A}}_s } \left( y \right)$ $= {1 \mathord{\left/ {\vphantom {1 2}} \right. \kern-\nulldelimiterspace} 2} + \left( {{1 \mathord{\left/ {\vphantom {1 2}} \right. \kern-\nulldelimiterspace} 2}} \right){\rm{erf}}\left( {{{\left( {\ln \left( y \right) - \mu _s } \right) } \mathord{\left/ {\vphantom {{\left( {\ln \left( y \right) - \mu _s } \right) } {\left( {\sqrt 2 \sigma _s } \right)}}} \right. \kern-\nulldelimiterspace} {\left( {\sqrt 2 \sigma _s } \right)}}} \right)$, $\bar m_{{\mathcal{A}}_s } \left( {\nu ,y} \right) = \exp \left\{ {\nu \mu _s  + \left( {{1 \mathord{\left/ {\vphantom {1 2}} \right. \kern-\nulldelimiterspace} 2}} \right)\nu ^2 \sigma _s^2 } \right\} - m_{{\rm{A}}_s } \left( {\nu ,y} \right)$, $\bar F_{{\mathcal{A}}_s } \left( y \right) = 1 - F_{{\mathcal{A}}_s } \left( y \right)$, $\mu _s  = \mu _s^{\left( {{\rm{dB}}} \right)} {{\ln \left( {10} \right)} \mathord{\left/ {\vphantom {{\ln \left( {10} \right)} {10}}} \right. \kern-\nulldelimiterspace} {10}}$ and $\sigma _s  = \sigma _s^{\left( {{\rm{dB}}} \right)} {{\ln \left( {10} \right)} \mathord{\left/ {\vphantom {{\ln \left( {10} \right)} {10}}} \right. \kern-\nulldelimiterspace} {10}}$.

\emph{Proof}: See Appendix I. \hfill $\Box$
\end{corollary}
\begin{remark} \label{Remark_2BallLogN}
Based on \textit{Lemma \ref{Intensity2Ball_Lemma}} and \textit{Corollary \ref{Intensity2Ball_Corollary}}, the intensity $\bar \Lambda _\Phi  \left( \cdot \right)$ in \eqref{Eq_11bis} can be expressed in closed-form as $\bar \Lambda _\Phi  \left( {\left[ {0,x} \right)} \right) \approx \bar \Lambda _\Phi ^{\left( {{\rm{approx}}} \right)} \left( {\left[ {0,x} \right)} \right) = \bar \Lambda _{{\rm{LOS}}}^{\left( {{\rm{approx}}} \right)} \left( {\left[ {0,x} \right)} \right) + \bar \Lambda _{{\rm{NLOS}}}^{\left( {{\rm{approx}}} \right)} \left( {\left[ {0,x} \right)} \right)$. This shows the usefulness of the two-ball approximation in the presence of a realistic channel model. Likewise, the approximation $\tilde \Lambda _\Phi  \left( {\left[ {0,x} \right)} \right) \approx \tilde \Lambda _\Phi ^{\left( {{\rm{approx}}} \right)} \left( {\left[ {0,x} \right)} \right) = \Lambda _{{\rm{LOS}}}^{\left( {{\rm{approx}}} \right)} \left( {\left[ {0,{\mathcal{A}}_{{\rm{LOS}}} x} \right)} \right) + \Lambda _{{\rm{NLOS}}}^{\left( {{\rm{approx}}} \right)} \left( {\left[ {0,{\mathcal{A}}_{{\rm{NLOS}}} x} \right)} \right)$ holds. Thus, the CDF of $\min \left\{ {\Phi ^{\left( {{\rm{approx}}} \right)} } \right\}$ follows from \textit{Lemma \ref{CDF_Lemma}} and the approximation $F_{\Phi ^{\left( 0 \right)} } \left( x \right) \approx 1 - \exp \left( { - \bar \Lambda _\Phi ^{\left( {{\rm{approx}}} \right)} \left( {\left[ {0,x} \right)} \right)} \right)$ holds. \hfill $\Box$
\end{remark}

We are now in the position of describing the procedure for computing the 15 parameters of the approximation in \eqref{Eq_2Ball__1}, \textit{i.e.}, $\left( {D_1 ,D_2 ,q_{\bar s}^{\left[ {0,D_1 } \right]} ,q_{\bar s}^{\left[ {D_1 ,D_2 } \right]} ,q_{\bar s}^{\left[ {D_2 ,\infty } \right]} } \right)$ for ${\bar s \in \left\{ {{\rm{LOS}},{\rm{NLOS}},{\rm{OUT}}} \right\}}$. Let the PPP of the path-loss $L = \left\{ {L_{{\rm{LOS}}} ,L_{{\rm{NLOS}}} ,L_{{\rm{OUT}}} } \right\}$ based on \eqref{Eq_4}. From \textit{Lemma \ref{Intensity_Lemma}} and \textit{Remark \ref{Remark_GeneralPPP}}, its intensity is $\Lambda _L \left( {\left[ {0,x} \right)} \right) = \left. {\tilde \Lambda _\Phi  \left( {\left[ {0,x} \right)} \right)} \right|_{{\mathcal{A}}_{{\rm{LOS}}}  = {\mathcal{A}}_{{\rm{NLOS}}}  = 1}  = \Lambda _{{\rm{LOS}}} \left( {\left[ {0,x} \right)} \right) + \Lambda _{{\rm{NLOS}}} \left( {\left[ {0,x} \right)} \right)$, where $\Lambda _{{\rm{LOS}}} \left(  \cdot  \right)$ and $\Lambda _{{\rm{NLOS}}} \left(  \cdot  \right)$ are defined in \eqref{Eq_11bis}. Let the PPP of the path-loss $L^{\left( {{\rm{approx}}} \right)}  = \left\{ {L_{{\rm{LOS}}}^{\left( {{\rm{approx}}} \right)} ,L_{{\rm{NLOS}}}^{\left( {{\rm{approx}}} \right)} ,L_{{\rm{OUT}}}^{\left( {{\rm{approx}}} \right)} } \right\}$ based on \eqref{Eq_2Ball__1}. From \textit{Lemma \ref{Intensity2Ball_Lemma}} and \textit{Remark \ref{Remark_2BallLogN}}, its intensity is $\Lambda _L^{\left( {{\rm{approx}}} \right)} \left( {\left[ {0,x} \right)} \right) = \left. {\tilde \Lambda _\Phi ^{\left( {{\rm{approx}}} \right)} \left( {\left[ {0,x} \right)} \right)} \right|_{{\mathcal{A}}_{{\rm{LOS}}}  = {\mathcal{A}}_{{\rm{NLOS}}}  = 1}  = \Lambda _{{\rm{LOS}}}^{\left( {{\rm{approx}}} \right)} \left( {\left[ {0,x} \right)} \right) + \Lambda _{{\rm{NLOS}}}^{\left( {{\rm{approx}}} \right)} \left( {\left[ {0,x} \right)} \right)$, where $\Lambda _{{\rm{LOS}}}^{\left( {{\rm{approx}}} \right)} \left(  \cdot  \right)$ and $\Lambda _{{\rm{NLOS}}}^{\left( {{\rm{approx}}} \right)} \left(  \cdot  \right)$ are in \eqref{Eq_2Ball__2}. The proposed matching procedure consists of two steps:
\begin{enumerate}
\item The first step lies in computing the 15 parameters in \eqref{Eq_2Ball__1} as the best fit of the \textit{unconstrained} optimization problem as follows ($x \in \left[ {0, + \infty } \right)$, $\bar s \in \left\{ {{\rm{LOS}},{\rm{NLOS}},{\rm{OUT}}} \right\}$):
\begin{equation}
\label{Eq_2Ball__4}
\mathop {\min }\limits_{\left( {D_1 ,D_2 ,q_{\bar s}^{\left[ {0,D_1 } \right]} ,q_{\bar s}^{\left[ {D_1 ,D_2 } \right]} ,q_{\bar s}^{\left[ {D_2 ,\infty } \right]} } \right)} \left\{ {\frac{1}{2}\left\| {\ln \left( {\Lambda _L \left( {\left[ {0,x} \right)} \right)} \right) - \ln \left( {\Lambda _L^{\left( {{\rm{approx}}} \right)} \left( {\left[ {0,x} \right)} \right)} \right)} \right\|_F^2 } \right\}
\end{equation}
\noindent where $\left\|  \cdot  \right\|_F$ denotes the Frobenius norm. The initial point for solving \eqref{Eq_2Ball__4} is randomly chosen. The optimization problem is unconstrained, since the approximation constraint in \eqref{Eq_2Ball__1} is neglected. The solution of \eqref{Eq_2Ball__4} is denoted by $\left( {\hat D_1 ,\hat D_2 ,\hat q_{\bar s}^{\left[ {0,D_1 } \right]} ,\hat q_{\bar s}^{\left[ {D_1 ,D_2 } \right]} ,\hat q_{\bar s}^{\left[ {D_2 ,\infty } \right]} } \right)$.
\item The second step lies in computing the 15 parameters in \eqref{Eq_2Ball__1} as the best fit of the \textit{constrained} optimization problem still formulated as in \eqref{Eq_2Ball__4}, but by taking into account the approximation constraint in \eqref{Eq_2Ball__1} and by assuming as the initial point of the search the solution of the first step, \textit{i.e.}, $\left( {\hat D_1 ,\hat D_2 ,\hat q_{\bar s}^{\left[ {0,D_1 } \right]} ,\hat q_{\bar s}^{\left[ {D_1 ,D_2 } \right]} ,\hat q_{\bar s}^{\left[ {D_2 ,\infty } \right]} } \right)$.
\end{enumerate}
\begin{remark} \label{Remark_2BallComments}
In practice, the unconstrained and constrained optimization problems can be solved by using the Matlab built-in functions \textsf{lsqcurvefit} and \textsf{fmincon}. The reason why a two-step approach is proposed is that we have found that solving first an unconstrained optimization problem provides results that are (almost) independent of the initial starting point of the search. The reason why the logarithm of the intensity instead of the intensity itself is matched is due to the possibility of better controlling the accuracy of the exponential functions in $\Lambda _L \left( \cdot \right)$. \hfill $\Box$
\end{remark}
\begin{table}[!t] \footnotesize
\renewcommand{\arraystretch}{1.5}
\caption{Three-state link and path-loss models from \cite[Table I]{Rappaport__mmWaveJSAC} and corresponding two-ball approximation obtained by using the algorithm described in Section \ref{TwoBall_Approximation}. The probabilities of being in an outage state are, by definition, $q_{{\rm{OUT}}}^{\left[ {0,D_1 } \right]} = 1 - q_{{\rm{LOS}}}^{\left[ {0,D_1 } \right]} - q_{{\rm{NLOS}}}^{\left[ {0,D_1 } \right]}$, $q_{{\rm{OUT}}}^{\left[ {D_1,D_2 } \right]} = 1 - q_{{\rm{LOS}}}^{\left[ {D_1,D_2 } \right]} - q_{{\rm{NLOS}}}^{\left[ {D_1,D_2 } \right]}$ and $q_{{\rm{OUT}}}^{\left[ {D_2, \infty} \right]} = 1 - q_{{\rm{LOS}}}^{\left[ {D_2,\infty } \right]} - q_{{\rm{NLOS}}}^{\left[ {D_2,\infty } \right]}$. \vspace{-0.5cm}} \label{Table_I}
\begin{center}
\begin{tabular}{|c|c|c|}
\hline
Carrier frequency ($F_c$) & Three-state link and path-loss models (\cite[Table I]{Rappaport__mmWaveJSAC}, \eqref{Eq_4}, \eqref{Eq_5}) & Two-ball approximation \\
\hline
\hline
 & $\alpha _{{\rm{LOS}}}  = 61.4$ dB, $\beta _{{\rm{LOS}}}  = 2$ & $D_1 = 56.9945$, $D_2 = 201.4371$ \\
28 GHz & $\alpha _{{\rm{NLOS}}}  = 72$ dB, $\beta _{{\rm{NLOS}}}  = 2.92$ & $q_{{\rm{LOS}}}^{\left[ {0,D_1 } \right]} = 0.8282$, $q_{{\rm{NLOS}}}^{\left[ {0,D_1 } \right]} = 0.1718$ \\
 & $\delta _{{\rm{LOS}}}  = 1/67.1$, $\gamma _{{\rm{LOS}}}  = 1$ & $q_{{\rm{LOS}}}^{\left[ {D_1,D_2 } \right]}= 0.1216$, $q_{{\rm{NLOS}}}^{\left[ {D_1,D_2 } \right]}= 0.7424$ \\
 & $\delta _{{\rm{OUT}}}  = 5.2$ $\gamma _{{\rm{OUT}}}  = 1/30$ & $q_{{\rm{LOS}}}^{\left[ {D_2 ,\infty } \right]} = 0$, $q_{{\rm{NLOS}}}^{\left[ {D_2 ,\infty } \right]} = 0$ \\
\hline
 & $\alpha _{{\rm{LOS}}}  = 69.8$ dB, $\beta _{{\rm{LOS}}}  = 2$ & $D_1 = 53.6287$, $D_2 = 195.3275$ \\
 73 GHz & $\alpha _{{\rm{NLOS}}}  = 82.7$ dB, $\beta _{{\rm{NLOS}}}  = 2.69$ & $q_{{\rm{LOS}}}^{\left[ {0,D_1 } \right]} = 0.8670$, $q_{{\rm{NLOS}}}^{\left[ {0,D_1 } \right]} = 0.1330$ \\
 & $\delta _{{\rm{LOS}}}  = 1/67.1$, $\gamma _{{\rm{LOS}}}  = 1$ & $q_{{\rm{LOS}}}^{\left[ {D_1,D_2 } \right]}= 0.1339$, $q_{{\rm{NLOS}}}^{\left[ {D_1,D_2 } \right]}= 0.7889$ \\
 & $\delta _{{\rm{OUT}}}  = 5.2$ $\gamma _{{\rm{OUT}}}  = 1/30$ & $q_{{\rm{LOS}}}^{\left[ {D_2 ,\infty } \right]} = 0$, $q_{{\rm{NLOS}}}^{\left[ {D_2 ,\infty } \right]} = 0$ \\
\hline
\end{tabular}
\end{center} \vspace{-1.25cm}
\end{table}
By applying the proposed two-step approximation technique to the empirical three-state link model proposed in \cite[Table I]{Rappaport__mmWaveJSAC}, the approximation in Table \ref{Table_I} is obtained. The accuracy of this approximation is studied in Section \ref{Results}. Besides being more mathematically tractable without loosing in accuracy, the two-ball approximation allows us to draw some interesting conclusions about the connectivity potential of mmWave communications. In particular:
\begin{enumerate}
  \item If the BS-to-MT distance $r$ is less than (about) 50 meters, \textit{i.e.}, $r<D_1$, we note that no link outage occurs. In other words, a link can be either in a LOS or a NLOS state. Furthermore, the probability of being in a LOS state is greater than 80\%.
  \item If the BS-to-MT distance $r$ is greater than (about) 50 meters but less than (about) 200 meters, \textit{i.e.}, $r \in \left[ {D_1 ,D_2 } \right]$, we note that a link can be in any of the three possible states. Furthermore, most likely, the MT is served by a NLOS BS.
  \item If the BS-to-MT distance $r$ is greater than (about) 200 meters, \textit{i.e.}, $r>D_2$, we note that the link is most likely to be in outage: no communication between BS and MT is possible.
  \item The distance $D_2$ identifies a critical operating regime, which is specific of mmWave communications and that it is not observed at $\mu$Wave communications that are characterized by a two-state link model. It is worth noting that $D_2$ is approximately equal to 200 meters, which is in agreement with the conclusions drawn in \cite{Rappaport__mmWaveAccess} and \cite{Rappaport__mmWaveJSAC}.
  \item The link state probabilities originating from the two-ball approximation in Table \ref{Table_I} provide useful guidelines on how to choose the average cell radius of mmWave systems. Radii of the order of 50 meters are expected to guarantee a very good connectivity, at the cost of a denser deployment. Radii larger than 200 meters, on the other hand, are expected to be too big for establishing a sufficiently reliable connection between BS and MT.
  \item Table \ref{Table_I} shows that the connectivity properties of mmWave networks operating at 28 GHz and 73 GHz are very similar. This is an important finding, since 28 GHz and 73 GHz represent the lower- and the upper-end, respectively, of the frequency range currently being considered for mmWave cellular communications.
\end{enumerate}
\subsection{Communication Blockage Probability} \label{Pr_OUT}
As mentioned in \textit{Remark \ref{Remark_Blockage}}, the peculiarity of the three-state link model in Section \ref{LinkStateModeling} lies is the presence of communication blockages if no BSs are available for serving the MT. The following lemma provides a closed-form expression of the probability that this event occurs.
\begin{lemma} \label{BlockageProbability_Lemma}
The probability ${\mathcal{P}}_{{\rm{blockage}}}  = \Pr \left\{ {\Psi _{{\rm{LOS}}}  = \emptyset  \cap \Psi _{{\rm{NLOS}}}  = \emptyset } \right\}$ that a communication blockage occurs can be formulated as ${\mathcal{P}}_{{\rm{blockage}}}  = \exp \left( { - \Lambda _{{\rm{blockage}}} } \right)$, where:
\begin{equation}
\label{Eq_12bis}
\Lambda _{{\rm{blockage}}}  = \pi \lambda \left( {\delta _{{\rm{OUT}}}^{ - 1} \ln \left( {\gamma _{{\rm{OUT}}} } \right)} \right)^2  + 2\pi \lambda \delta _{{\rm{OUT}}}^{ - 2} \gamma _{{\rm{OUT}}} \left( {\gamma _{{\rm{OUT}}}^{ - 1}  + \gamma _{{\rm{OUT}}}^{ - 1} \ln \left( {\gamma _{{\rm{OUT}}} } \right)} \right)
\end{equation}

\emph{Proof}: See Appendix I. \hfill $\Box$
\end{lemma}
\begin{remark} \label{Remark_CommunBlockage}
The communication blockage probability in \eqref{Eq_12bis} is independent of the cell association criterion. Also, ${\mathcal{P}}_{{\rm{blockage}}} = 0$ if $\delta _{{\rm{OUT}}}  = 0$, \textit{i.e.}, $p_{{\rm{OUT}}} \left( r \right) = 0$ in \eqref{Eq_4}. In general, thus, let ${\mathcal{P}}_{{\rm{LOS}}}$ and ${\mathcal{P}}_{{\rm{NLOS}}}$ be the probabilities that the MT is served by a LOS and a NLOS BS, respectively, we have ${\mathcal{P}}_{{\rm{LOS}}}  + {\mathcal{P}}_{{\rm{NLOS}}} + {\mathcal{P}}_{{\rm{blockage}}} = 1$ and ${\mathcal{P}}_{{\rm{LOS}}}  + {\mathcal{P}}_{{\rm{NLOS}}}  \le 1$. This implies that the coverage probability may be zero even for ${\rm{T}} = 0$. This occurs if ${\mathcal{P}}_{{\rm{LOS}}}  = {\mathcal{P}}_{{\rm{NLOS}}}  = 0$ and ${\mathcal{P}}_{{\rm{blockage}}} = 1$. A similar comment applies to the average rate. By direct inspection of \eqref{Eq_12bis}, this occurs if $\delta _{{\rm{OUT}}}  \to  + \infty$, which corresponds to $p_{{\rm{LOS}}} \left( r \right) = p_{{\rm{NLOS}}} \left( r \right) = 0$ and $p_{{\rm{OUT}}} \left( r \right) = 1$ in \eqref{Eq_4}. As discussed in Section \ref{TwoBall_Approximation}, there is a critical distance where this operating regime emerges, which corresponds to 200 meters for the considered mmWave channel model. \hfill $\Box$
\end{remark}
\section{Modeling Coverage Probability and Average Rate} \label{CoverageRate}
\subsection{Smallest Path-Loss Cell Association} \label{CoverageRate_PathLoss}
Assume a cell association based on the smallest path-loss and no beamsteering errors. From \eqref{Eq_6}, $U^{\left( 0 \right)}  = {{\mathsf{P}G^{\left( 0 \right)} \left| {h_{s}^{\left( {\rm{0}} \right)} } \right|^2 } \mathord{\left/ {\vphantom {{\mathsf{P}G^{\left( 0 \right)} \left| {h_{ s}^{\left( {\rm{0}} \right)} } \right|^2 } {L^{\left( 0 \right)} }}} \right. \kern-\nulldelimiterspace} {L^{\left( 0 \right)} }}$, where $s=\rm{LOS}$ or $s=\rm{NLOS}$ if the MT is served by a LOS or a NLOS BS, respectively, and $U^{\left( 0 \right)} = 0$ if a communication blockage occurs. Then, the SNR can be formulated as follows:
\begin{equation}
\label{Eq_1__SecIV}
{\rm{SNR}}\mathop  = \limits^{\left( a \right)} \frac{{\mathsf{P}G^{\left( 0 \right)} \left| {h_{{\rm{LOS}}}^{\left( {\rm{0}} \right)} } \right|^2 }}{{\sigma _N^2 L^{\left( 0 \right)} }}\delta \left\{ {L^{\left( 0 \right)}  - L_{{\rm{LOS}}}^{\left( 0 \right)} } \right\} + \frac{{\mathsf{P}G^{\left( 0 \right)} \left| {h_{{\rm{NLOS}}}^{\left( {\rm{0}} \right)} } \right|^2 }}{{\sigma _N^2 L^{\left( 0 \right)} }}\delta \left\{ {L^{\left( 0 \right)}  - L_{{\rm{NLOS}}}^{\left( 0 \right)} } \right\}
\end{equation}
\noindent where (a) takes into account that the distribution of LOS and NLOS links is different.
\begin{proposition} \label{Pcov_Prop}
Let the SNR in \eqref{Eq_1__SecIV}. The coverage probability in \eqref{Eq_8} can be formulated as follows:
\begin{equation}
\label{Eq_2__SecIV}
\begin{split}
 & {\rm{P}}^{\left( {{\mathop{\rm cov}} } \right)} \left( {\rm{T}} \right) = {\rm{P}}_{{\rm{LOS}}}^{\left( {{\mathop{\rm cov}} } \right)} \left( {\rm{T}} \right) + {\rm{P}}_{{\rm{NLOS}}}^{\left( {{\mathop{\rm cov}} } \right)} \left( {\rm{T}} \right) \\
 & {\rm{P}}_s^{\left( {{\mathop{\rm cov}} } \right)} \left( {\rm{T}} \right) = \frac{1}{2}\int\nolimits_0^{ + \infty } {{\rm{erfc}}\left( {\frac{{{\rm{ln}}\left( {{{{\rm{T}}x} \mathord{\left/
 {\vphantom {{{\rm{T}}x} {\gamma ^{\left( 0 \right)} }}} \right.
 \kern-\nulldelimiterspace} {\gamma ^{\left( 0 \right)} }}} \right) - \mu _s }}{{\sqrt 2 \sigma _s }}} \right)\Lambda _{L_s }^{\left( 1 \right)} \left( {\left[ {0,x} \right)} \right)\exp \left( { - \Lambda _L \left( {\left[ {0,x} \right)} \right)} \right)dx}  \\
 \end{split}
\end{equation}
\noindent where $s = \left\{ {{\rm{LOS}},{\rm{NLOS}}} \right\}$, $\gamma ^{\left( 0 \right)}  = {{{\mathsf{P}}G^{\left( 0 \right)} } \mathord{\left/ {\vphantom {{{\mathsf{P}}G^{\left( 0 \right)} } {\sigma _N^2 }}} \right. \kern-\nulldelimiterspace} {\sigma _N^2 }}$, $\mu _s$ and $\sigma _s$ are defined in \textit{Corollary \ref{Intensity2Ball_Corollary}}, $\Lambda _L \left( {\left[ {0,x} \right)} \right) = \Lambda _{{\rm{LOS}}} \left( {\left[ {0,x} \right)} \right) + \Lambda _{{\rm{NLOS}}} \left( {\left[ {0,x} \right)} \right)$, where $\Lambda _{{\rm{LOS}}} \left(  \cdot  \right)$ and $\Lambda _{{\rm{NLOS}}} \left(  \cdot  \right)$ are defined in \textit{Lemma \ref{Intensity_Lemma}}, $\Lambda _{L_{{\rm{LOS}}} }^{\left( 1 \right)} \left( {\left[ {0,x} \right)} \right) = \Upsilon _0^{\left( 1 \right)} \left( {x;{\rm{LOS}}} \right)$, where $\Upsilon _0^{\left( 1 \right)} \left( { \cdot ; \cdot } \right)$ is the first derivative of $\Upsilon _0 \left( { \cdot ; \cdot } \right)$ defined in \textit{Lemma \ref{Intensity_Lemma}}, \textit{i.e.}, $\Upsilon _0^{\left( 1 \right)} \left( {x;s} \right) = {{d\Upsilon _0 \left( {x;s} \right)} \mathord{\left/ {\vphantom {{d\Upsilon _0 \left( {x;s} \right)} {dx}}} \right. \kern-\nulldelimiterspace} {dx}}$, $\Lambda _{L_{{\rm{NLOS}}} }^{\left( 1 \right)} \left( {\left[ {0,x} \right)} \right) = \Upsilon _1^{\left( 1 \right)} \left( {x;{\rm{NLOS}}} \right) - \Upsilon _0^{\left( 1 \right)} \left( {x;{\rm{NLOS}}} \right)$, where $\Upsilon _1^{\left( 1 \right)} \left( { \cdot ; \cdot } \right)$ is the first derivative of $\Upsilon _1 \left( { \cdot ; \cdot } \right)$ defined in \textit{Lemma \ref{Intensity_Lemma}}, \textit{i.e.}, $\Upsilon _1^{\left( 1 \right)} \left( {x;s} \right) = {{d\Upsilon _1 \left( {x;s} \right)} \mathord{\left/ {\vphantom {{d\Upsilon _1 \left( {x;s} \right)} {dx}}} \right. \kern-\nulldelimiterspace} {dx}}$. $\Upsilon _0^{\left( 1 \right)} \left( { \cdot ; \cdot } \right)$ and $\Upsilon _1^{\left( 1 \right)} \left( { \cdot ; \cdot } \right)$ can be formulated as follows:
\begin{equation}
\label{Eq_3__SecIV}
\begin{split}
 & \Upsilon _0^{\left( 1 \right)} \left( {x;s} \right) = {\mathcal{K}}_2 \left( {e^{ - W}  + We^{ - W}  - e^{ - V_s x^{{1 \mathord{\left/
 {\vphantom {1 {\beta _s }}} \right.
 \kern-\nulldelimiterspace} {\beta _s }}} }  - V_s x^{{1 \mathord{\left/
 {\vphantom {1 {\beta _s }}} \right.
 \kern-\nulldelimiterspace} {\beta _s }}} e^{ - V_s x^{{1 \mathord{\left/
 {\vphantom {1 {\beta _s }}} \right.
 \kern-\nulldelimiterspace} {\beta _s }}} } } \right)\delta \left( {x - Z_s } \right) \\
 & - {\mathcal{K}}_1 \left( {1 - e^{ - Q_s x^{{1 \mathord{\left/
 {\vphantom {1 {\beta _s }}} \right.
 \kern-\nulldelimiterspace} {\beta _s }}} }  - Q_s x^{{1 \mathord{\left/
 {\vphantom {1 {\beta _s }}} \right.
 \kern-\nulldelimiterspace} {\beta _s }}} e^{ - Q_s x^{{1 \mathord{\left/
 {\vphantom {1 {\beta _s }}} \right.
 \kern-\nulldelimiterspace} {\beta _s }}} } } \right)\delta \left( {x - Z_s } \right)+ {\mathcal{K}}_1 \left( {1 - e^{ - R}  - Re^{ - R} } \right)\delta \left( {x - Z_s } \right) \\
 & + {\mathcal{K}}_2 \left( {{{V_s^2 } \mathord{\left/
 {\vphantom {{V_s^2 } {\beta _s }}} \right.
 \kern-\nulldelimiterspace} {\beta _s }}} \right)x^{{2 \mathord{\left/
 {\vphantom {2 {\beta _s }}} \right.
 \kern-\nulldelimiterspace} {\beta _s }} - 1} e^{ - V_s x^{{1 \mathord{\left/
 {\vphantom {1 {\beta _s }}} \right.
 \kern-\nulldelimiterspace} {\beta _s }}} } {\mathcal{H}}\left( {x - Z_s } \right) + {\mathcal{K}}_1 \left( {{{Q_s^2 } \mathord{\left/
 {\vphantom {{Q_s^2 } {\beta _s }}} \right.
 \kern-\nulldelimiterspace} {\beta _s }}} \right)x^{{2 \mathord{\left/
 {\vphantom {2 {\beta _s }}} \right.
 \kern-\nulldelimiterspace} {\beta _s }} - 1} e^{ - Q_s x^{{1 \mathord{\left/
 {\vphantom {1 {\beta _s }}} \right.
 \kern-\nulldelimiterspace} {\beta _s }}} } {\mathcal{\bar H}}\left( {x - Z_s } \right) \\
 & \mathop  = \limits^{\left( a \right)} {\mathcal{K}}_2 \left( {{{V_s^2 } \mathord{\left/
 {\vphantom {{V_s^2 } {\beta _s }}} \right.
 \kern-\nulldelimiterspace} {\beta _s }}} \right)x^{{2 \mathord{\left/
 {\vphantom {2 {\beta _s }}} \right.
 \kern-\nulldelimiterspace} {\beta _s }} - 1} e^{ - V_s x^{{1 \mathord{\left/
 {\vphantom {1 {\beta _s }}} \right.
 \kern-\nulldelimiterspace} {\beta _s }}} } {\mathcal{H}}\left( {x - Z_s } \right) + {\mathcal{K}}_1 \left( {{{Q_s^2 } \mathord{\left/
 {\vphantom {{Q_s^2 } {\beta _s }}} \right.
 \kern-\nulldelimiterspace} {\beta _s }}} \right)x^{{2 \mathord{\left/
 {\vphantom {2 {\beta _s }}} \right.
 \kern-\nulldelimiterspace} {\beta _s }} - 1} e^{ - Q_s x^{{1 \mathord{\left/
 {\vphantom {1 {\beta _s }}} \right.
 \kern-\nulldelimiterspace} {\beta _s }}} } {\mathcal{\bar H}}\left( {x - Z_s } \right) \\
\end{split}
\end{equation}
\begin{equation}
\label{Eq_4__SecIV}
\begin{split}
 & \Upsilon _1^{\left( 1 \right)} \left( {x;s} \right) =  - \pi \lambda \kappa _s^{ - 2} x^{{2 \mathord{\left/
 {\vphantom {2 {\beta _s }}} \right.
 \kern-\nulldelimiterspace} {\beta _s }}} \delta \left( {x - Z_s } \right) + \pi \lambda \left( {\delta _{{\rm{OUT}}}^{ - 1} \ln \left( {\gamma _{{\rm{OUT}}} } \right)} \right)^2 \delta \left( {x - Z_s } \right) \\
 & + 2\pi \lambda \delta _{{\rm{OUT}}}^{ - 2} \gamma _{{\rm{OUT}}} \left( {\gamma _{{\rm{OUT}}}^{ - 1}  + \gamma _{{\rm{OUT}}}^{ - 1} \ln \left( {\gamma _{{\rm{OUT}}} } \right) - e^{ - T_s x^{{1 \mathord{\left/
 {\vphantom {1 {\beta _s }}} \right.
 \kern-\nulldelimiterspace} {\beta _s }}} }  - T_s x^{{1 \mathord{\left/
 {\vphantom {1 {\beta _s }}} \right.
 \kern-\nulldelimiterspace} {\beta _s }}} e^{ - T_s x^{{1 \mathord{\left/
 {\vphantom {1 {\beta _s }}} \right.
 \kern-\nulldelimiterspace} {\beta _s }}} } } \right)\delta \left( {x - Z_s } \right) \\
 & + 2\pi \lambda \kappa _s^{ - 2} \beta _s^{ - 1} x^{{2 \mathord{\left/
 {\vphantom {2 {\beta _s }}} \right.
 \kern-\nulldelimiterspace} {\beta _s }} - 1} {\mathcal{\bar H}}\left( {x - Z_s } \right) + 2\pi \lambda \delta _{{\rm{OUT}}}^{ - 2} \gamma _{{\rm{OUT}}} T_s^2 \beta _s^{ - 1} x^{{2 \mathord{\left/
 {\vphantom {2 {\beta _s }}} \right.
 \kern-\nulldelimiterspace} {\beta _s }} - 1} e^{ - T_s x^{{1 \mathord{\left/
 {\vphantom {1 {\beta _s }}} \right.
 \kern-\nulldelimiterspace} {\beta _s }}} } {\mathcal{H}}\left( {x - Z_s } \right) \\
 & \mathop  = \limits^{\left( b \right)} 2\pi \lambda \kappa _s^{ - 2} \beta _s^{ - 1} x^{{2 \mathord{\left/
 {\vphantom {2 {\beta _s }}} \right.
 \kern-\nulldelimiterspace} {\beta _s }} - 1} {\mathcal{\bar H}}\left( {x - Z_s } \right) + 2\pi \lambda \delta _{{\rm{OUT}}}^{ - 2} \gamma _{{\rm{OUT}}} T_s^2 \beta _s^{ - 1} x^{{2 \mathord{\left/
 {\vphantom {2 {\beta _s }}} \right.
 \kern-\nulldelimiterspace} {\beta _s }} - 1} e^{ - T_s x^{{1 \mathord{\left/
 {\vphantom {1 {\beta _s }}} \right.
 \kern-\nulldelimiterspace} {\beta _s }}} } {\mathcal{H}}\left( {x - Z_s } \right) \\
\end{split}
\end{equation}

\emph{Proof}: See Appendix II. \hfill $\Box$
\end{proposition}

\textit{Proposition \ref{Pcov_Prop}} provides an exact single-integral expression of the coverage probability. In particular, the two-ball approximation in Section \ref{TwoBall_Approximation}, which may be obtained by replacing $\Lambda _{{\rm{LOS}}} \left(  \cdot  \right)$ and $\Lambda _{{\rm{NLOS}}} \left(  \cdot  \right)$ with $\Lambda _{{\rm{LOS}}}^{\left( {{\rm{approx}}} \right)} \left(  \cdot  \right)$ and $\Lambda _{{\rm{NLOS}}}^{\left( {{\rm{approx}}} \right)} \left(  \cdot  \right)$ in \eqref{Eq_2Ball__2}, respectively, is not used. The average rate can be computed from \eqref{Eq_9}, \textit{e.g.}, by using the GCQ formulation. In the most general setup considered in this paper, \eqref{Eq_2__SecIV} cannot be further simplified, even using the two-ball approximation. A simplified formulation can be obtained, however, in some special cases.
\begin{corollary} \label{Pcov_Cor}
Let the SNR in \eqref{Eq_1__SecIV} under the assumption that ${\left| {h_{{\rm{LOS}}}^{\left( 0 \right)} } \right|}$ and ${\left| {h_{{\rm{NLOS}}}^{\left( 0 \right)} } \right|}$ are independent and identically distributed, \textit{i.e.}, $\mu^{\left( {{\rm{dB}}} \right)} = \mu _{{\rm{LOS}}}^{\left( {{\rm{dB}}} \right)}  = \mu _{{\rm{NLOS}}}^{\left( {{\rm{dB}}} \right)}$ and $\sigma^{\left( {{\rm{dB}}} \right)} = \sigma _{{\rm{LOS}}}^{\left( {{\rm{dB}}} \right)}  = \sigma _{{\rm{NLOS}}}^{\left( {{\rm{dB}}} \right)}$. The coverage probability in \eqref{Eq_8} can be formulated as follows:
\begin{equation}
\label{Eq_5__SecIV}
{\rm{P}}^{\left( {{\mathop{\rm cov}} } \right)} \left( {\rm{T}} \right) = \int\nolimits_0^{ + \infty } {F_{L^{\left( 0 \right)} } \left( {\frac{{\mathsf{P}G^{\left( 0 \right)} }}{{\sigma _N^2 {\rm{T}}}}x} \right)f_{\left| {h^{\left( {\rm{0}} \right)} } \right|^2 } \left( x \right)dx}
\end{equation}
\noindent where $F_{L^{\left( 0 \right)} } \left( \cdot \right)$ follows from \eqref{Eq_11cdf} with $L^{\left( 0 \right)}  = \left. {\Phi ^{\left( 0 \right)} } \right|_{{\mathcal{A}}_{{\rm{LOS}}}  = {\mathcal{A}}_{{\rm{NLOS}}}  = 1}$, $f_{\left| {h^{\left( 0 \right)} } \right|^2 } \left( \xi  \right) = f_{\left| {h_{{\rm{LOS}}}^{\left( 0 \right)} } \right|^2 } \left( \xi  \right) = f_{\left| {h_{{\rm{NLOS}}}^{\left( 0 \right)} } \right|^2 } \left( \xi  \right) = \left( {\sqrt {2\pi } \sigma x} \right)^{ - 1} \exp \left( { - {{\left( {\ln \left( x \right) - \mu } \right)^2 } \mathord{\left/{\vphantom {{\left( {\ln \left( x \right) - \mu} \right)^2 } {2\sigma^2 }}} \right. \kern-\nulldelimiterspace} {2\sigma^2 }}} \right)$ is the PDF of ${\left| {h_{{\rm{LOS}}}^{\left( 0 \right)} } \right|^2 }$ and ${\left| {h_{{\rm{NLOS}}}^{\left( 0 \right)} } \right|^2 }$ with $\mu  = \mu^{\left( {{\rm{dB}}} \right)} \left( {{{\ln \left( {10} \right)} \mathord{\left/ {\vphantom {{\ln \left( {10} \right)} {10}}} \right.  \kern-\nulldelimiterspace} {10}}} \right)$ and $\sigma  = \sigma^{\left( {{\rm{dB}}} \right)} \left( {{{\ln \left( {10} \right)} \mathord{\left/ {\vphantom {{\ln \left( {10} \right)} {10}}} \right.  \kern-\nulldelimiterspace} {10}}} \right)$.

\emph{Proof}: See Appendix II. \hfill $\Box$
\end{corollary}
\begin{remark} \label{Remark_PcovSameFading}
In \textit{Corollary \ref{Pcov_Cor}}, only the fading parameters of LOS and NLOS channels are assumed to be the same. The path-loss model and the link state probability of LOS and NLOS links are, on the other hand, still different and formulated in a general manner. Similar to \eqref{Eq_2__SecIV}, \eqref{Eq_5__SecIV} is still formulated in an integral form. The latter mathematical formulation has, however, two main advantages: i) it is simpler to be computed numerically and ii) it is provided in a general form that is applicable to any distributions, \textit{i.e.}, $f_{\left| {h^{\left( 0 \right)} } \right|^2 } \left(  \cdot  \right)$, of the fading power gains. \hfill $\Box$
\end{remark}
\begin{proposition} \label{Pcov_PropApprox}
Let the SNR in \eqref{Eq_1__SecIV} under the assumption that $\gamma ^{\left( 0 \right)}  = {{{\mathsf{P}}G^{\left( 0 \right)} } \mathord{\left/ {\vphantom {{{\mathsf{P}}G^{\left( 0 \right)} } {\sigma _N^2 }}} \right. \kern-\nulldelimiterspace} {\sigma _N^2 }} \gg 1$. The average rate in \eqref{Eq_9} can be formulated as ${\mathop{\rm R}\nolimits}  \approx {\rm{R}}_{{\rm{LOS}}}  + {\rm{R}}_{{\rm{NLOS}}}$, where ($s = \left\{ {{\rm{LOS}},{\rm{NLOS}}} \right\}$):
\begin{equation}
\label{Eq_6__SecIV}
 \begin{split}
 \hspace{-0.5cm} & {\mathop{\mathcal{ J}}\nolimits}_s \left( x \right) = \frac{1}{{\sqrt \pi  }}\exp \left( { - \frac{{\left( {\ln \left( {{x \mathord{\left/
 {\vphantom {x {\gamma ^{\left( 0 \right)} }}} \right.
 \kern-\nulldelimiterspace} {\gamma ^{\left( 0 \right)} }}} \right) - \mu _s } \right)^2 }}{{2\sigma _s^2 }}} \right) - \left( {\ln \left( {{x \mathord{\left/
 {\vphantom {x {\gamma ^{\left( 0 \right)} }}} \right.
 \kern-\nulldelimiterspace} {\gamma ^{\left( 0 \right)} }}} \right) - \mu _s } \right){\rm{erfc}}\left( {\frac{{\ln \left( {{x \mathord{\left/
 {\vphantom {x {\gamma ^{\left( 0 \right)} }}} \right.
 \kern-\nulldelimiterspace} {\gamma ^{\left( 0 \right)} }}} \right) - \mu _s }}{{\sqrt 2 \sigma _s }}} \right) \\
 \hspace{-0.5cm} & {\rm{R}}_s  = \left( {{1 \mathord{\left/
 {\vphantom {1 2}} \right.
 \kern-\nulldelimiterspace} 2}} \right)\int\nolimits_0^{ + \infty } {{\mathop{\mathcal{ J}}\nolimits}_s \left( x \right)\Lambda _{L_s }^{\left( 1 \right)} \left( {\left[ {0,x} \right)} \right)\exp \left( { - \Lambda _L \left( {\left[ {0,x} \right)} \right)} \right)dx}
 \end{split}
\end{equation}

\emph{Proof}: See Appendix II. \hfill $\Box$
\end{proposition}
\begin{remark} \label{Remark_ApproximationAccuracy}
The approximation in \eqref{Eq_6__SecIV} provides a single-integral expression of the average rate. Its accuracy is expected to increase as the directivity gain, $G^{(0)}$, of the intended link and the density, $\lambda$, of the BSs increase. In this case, in fact, ${\rm{SINR}} \approx {\rm{SNR}} \gg {\rm{1}}$ (see proof). \hfill $\Box$
\end{remark}
\begin{remark} \label{Remark_Trends}
By direct inspection of, \textit{e.g.}, \eqref{Eq_5__SecIV}, it follows that coverage probability and average rate increase as $\mathsf{P}$, ${G^{\left( 0 \right)} }$ and $\lambda$ increase. They decrease, on the other hand, as ${\sigma _N^2 }$ increases. Therefore, the performance of mmWave cellular networks improves by increasing the transmit power, the directivity gain of the intended link and the density of BSs. \hfill $\Box$
\end{remark}
\begin{remark} \label{Remark_NoBlocking}
From \textit{Lemmas \ref{Intensity_Lemma} \textnormal{and} \ref{CDF_Lemma}}, it follows that \textit{Proposition \ref{Pcov_Prop}}, \textit{Corollary \ref{Pcov_Cor}} and \textit{Proposition \ref{Pcov_PropApprox}} still hold if $p_{{\rm{OUT}}} \left( r \right) = 0$ in \eqref{Eq_4}. $\Upsilon _0 \left( {\cdot;\cdot} \right)$ and $\Upsilon _1 \left( {\cdot;\cdot} \right)$ in \textit{Corollary \ref{CDF_Corollary}} can be used. \hfill $\Box$
\end{remark}
\subsection{Highest Received Power Cell Association} \label{CoverageRate_Best}
Assume a cell association based on the highest received power and no beamsteering errors. From \eqref{Eq_7}, $U^{\left( 0 \right)}  = {{\mathsf{P}G^{\left( 0 \right)} } \mathord{\left/ {\vphantom {{\mathsf{P}G^{\left( 0 \right)} } {P^{\left( 0 \right)} }}} \right. \kern-\nulldelimiterspace} {P^{\left( 0 \right)} }}$ and the SNR can be written as ${\rm{SNR = }}{{\mathsf{P}G^{\left( 0 \right)} } \mathord{\left/{\vphantom {{\mathsf{P}G^{\left( 0 \right)} } {\left( {\sigma _N^2 P^{\left( 0 \right)} } \right)}}} \right. \kern-\nulldelimiterspace} {\left( {\sigma _N^2 P^{\left( 0 \right)} } \right)}}$. In particular, ${\rm{SNR}} = 0$ if a communication blockage occurs.
\begin{proposition} \label{PcovBest_Prop}
Let ${\rm{SNR = }}{{\mathsf{P}G^{\left( 0 \right)} } \mathord{\left/{\vphantom {{\mathsf{P}G^{\left( 0 \right)} } {\left( {\sigma _N^2 P^{\left( 0 \right)} } \right)}}} \right. \kern-\nulldelimiterspace} {\left( {\sigma _N^2 P^{\left( 0 \right)} } \right)}}$. The coverage probability in \eqref{Eq_8} is equal to:
\begin{equation}
\label{Eq_7__SecIV}
\begin{split}
 {\rm{P}}^{\left( {{\mathop{\rm cov}} } \right)} \left( {\rm{T}} \right) &= F_{P^{\left( 0 \right)} } \left( {{{{\mathsf{P}}G^{\left( 0 \right)} } \mathord{\left/
 {\vphantom {{{\mathsf{P}}G^{\left( 0 \right)} } {\left( {\sigma _N^2 {\rm{T}}} \right)}}} \right.
 \kern-\nulldelimiterspace} {\left( {\sigma _N^2 {\rm{T}}} \right)}}} \right)\mathop  = \limits^{\left( a \right)} 1 - \exp \left( {\Lambda _P \left( {\left[ {0,{{{\mathsf{P}}G^{\left( 0 \right)} } \mathord{\left/
 {\vphantom {{{\mathsf{P}}G^{\left( 0 \right)} } {\left( {\sigma _N^2 {\rm{T}}} \right)}}} \right.
 \kern-\nulldelimiterspace} {\left( {\sigma _N^2 {\rm{T}}} \right)}}} \right)} \right)} \right) \\
 & \mathop  \approx \limits^{\left( b \right)} 1 - \exp \left( {\Lambda _P^{\left( {{\rm{approx}}} \right)} \left( {\left[ {0,{{{\mathsf{P}}G^{\left( 0 \right)} } \mathord{\left/
 {\vphantom {{{\mathsf{P}}G^{\left( 0 \right)} } {\left( {\sigma _N^2 {\rm{T}}} \right)}}} \right.
 \kern-\nulldelimiterspace} {\left( {\sigma _N^2 {\rm{T}}} \right)}}} \right)} \right)} \right) \\
 \end{split}
\end{equation}
\noindent where $\Lambda _P \left( {\left[ {0,x} \right)} \right) = \bar \Lambda _\Phi  \left( {\left[ {0,x} \right)} \right)$ and $\Lambda _P^{\left( {{\rm{approx}}} \right)} \left( {\left[ {0,x} \right)} \right) = \bar \Lambda _\Phi ^{\left( {{\rm{approx}}} \right)} \left( {\left[ {0,x} \right)} \right)$ are defined in \eqref{Eq_11bis} and \textit{Remark \ref{Remark_2BallLogN}}, respectively.

\emph{Proof}: (a) follows from \textit{Lemma \ref{Intensity_Lemma}} by noting that $P^{\left( 0 \right)}  = \Phi ^{\left( 0 \right)}$ if ${{\mathcal{A}}_{{\rm{LOS}}} }$ and ${{\mathcal{A}}_{{\rm{NLOS}}} }$ are random variables and (b) follows from the two-ball approximation in Section \ref{TwoBall_Approximation}. \hfill $\Box$
\end{proposition}
\begin{remark} \label{Remark_BestAssocComment}
The equality in (a) provides an exact single-integral expression of the coverage probability. Hence, the average rate in \eqref{Eq_9} is formulated in terms of a two-fold integral. The two-ball approximation in (b), on the other hand, provides an approximated closed-form expression of the coverage probability, which results in a single-integral expression of the average rate. \hfill $\Box$
\end{remark}
\begin{remark} \label{Remark_Comparison}
Denote the coverage probabilities in \eqref{Eq_5__SecIV} and \eqref{Eq_7__SecIV} by ${\rm{P}}^{\left( {{\mathop{\rm cov}} {\rm{,  \, path - loss}}} \right)} \left( \cdot \right)$ and ${\rm{P}}^{\left( {{\mathop{\rm cov}} {\rm{,  \, power}}} \right)} \left( \cdot \right)$, respectively. With the aid of the Jensen inequality applied to the exponential function, it follows, as expected, that ${\rm{P}}^{\left( {{\mathop{\rm cov}} {\rm{,  \, path - loss}}} \right)} \left( {\rm{T}} \right) \le {\rm{P}}^{\left( {{\mathop{\rm cov}} {\rm{,  \, power}}} \right)} \left( {\rm{T}} \right)$. The price to be paid for the better performance provided by the cell association based on the highest received power is the need of knowing the instantaneous shadowing power gains. A similar performance trend is expected to hold when LOS and NLOS links have different distributions. The proof of this trend is, however, not straightforward from \eqref{Eq_2__SecIV}. As for the performance trends that originate from \eqref{Eq_7__SecIV}, the same comments as in \textit{Remark \ref{Remark_Trends}} apply. \hfill $\Box$
\end{remark}
\section{Generalizations} \label{Generalizations}
In Section \ref{CoverageRate}, new frameworks for computing coverage and rate of mmWave systems are provided, under the assumptions of no beamsteering errors and a single tier of BSs. With these assumptions, coverage and rate are formulated in an exact single and two-fold integral expressions, respectively. A closed-form expression of the coverage is provided for high-SNR and by assuming a cell association based on the smallest path-loss. A closed-form expression of the coverage is provided by relying on a two-ball approximation for modeling the link state and by assuming a cell association based on the highest received power. In this section, the assumptions of Section \ref{CoverageRate} are removed, without increasing the complexity of the frameworks.
\subsection{Beamforming Alignment Errors} \label{BeamformingErrors}
By direct inspection of the frameworks for computing coverage probability and average rate in Section \ref{CoverageRate}, it is apparent that they depend on $G^{\left( 0 \right)}  = G_{{\rm{BS}}}^{\left( 0 \right)} G_{{\rm{MT}}}^{\left( 0 \right)}$. In mathematical terms, this dependency can be highlighted by using the notation ${\rm{P}}^{\left( {{\mathop{\rm cov}} } \right)} \left( {{\rm{T}};G^{\left( 0 \right)} } \right)$ and ${\mathop{\rm R}\nolimits} \left( {G^{\left( 0 \right)} } \right)$.
\begin{proposition} \label{BeamError_Prop}
Let ${\rm{P}}^{\left( {{\mathop{\rm cov}} } \right)} \left( \cdot;G^{\left( 0 \right)} \right)$ and ${\mathop{\rm R}\nolimits} \left( {G^{\left( 0 \right)} } \right)$ be coverage probability and average rate, respectively, available in Section \ref{CoverageRate} for cell associations based on the smallest path-loss and the highest received power. Let the beamforming alignment error model in Section \ref{PointingErrorModeling}. The coverage probability can be formulated as follows:
\begin{equation}
\label{Eq_1__SecV}
\begin{split}
 {\rm{P}}^{\left( {{\mathop{\rm cov}} } \right)} \left( {\rm{T}} \right) &= {\mathbb{E}}_{G^{\left( 0 \right)} } \left\{ {{\rm{P}}^{\left( {{\mathop{\rm cov}} } \right)} \left( {{\rm{T}};G^{\left( 0 \right)} } \right)} \right\} = \int\nolimits_0^{ + \infty } {{\rm{P}}^{\left( {{\mathop{\rm cov}} } \right)} \left( {{\rm{T}};g} \right)f_{G^{\left( 0 \right)} } \left( g \right)dg}  \\
 & = F_{\left| {\varepsilon _{{\rm{BS}}} } \right|} \left( {{{\omega _{{\rm{BS}}} } \mathord{\left/
 {\vphantom {{\omega _{{\rm{BS}}} } 2}} \right.
 \kern-\nulldelimiterspace} 2}} \right)F_{\left| {\varepsilon _{{\rm{MT}}} } \right|} \left( {{{\omega _{{\rm{MT}}} } \mathord{\left/
 {\vphantom {{\omega _{{\rm{MT}}} } 2}} \right.
 \kern-\nulldelimiterspace} 2}} \right){\rm{P}}^{\left( {{\mathop{\rm cov}} } \right)} \left( {{\rm{T}};G_{{\rm{BS}}}^{\left( {\max } \right)} G_{{\rm{MT}}}^{\left( {\max } \right)} } \right) \\ & + F_{\left| {\varepsilon _{{\rm{BS}}} } \right|} \left( {{{\omega _{{\rm{BS}}} } \mathord{\left/
 {\vphantom {{\omega _{{\rm{BS}}} } 2}} \right.
 \kern-\nulldelimiterspace} 2}} \right)\bar F_{\left| {\varepsilon _{{\rm{MT}}} } \right|} \left( {{{\omega _{{\rm{MT}}} } \mathord{\left/
 {\vphantom {{\omega _{{\rm{MT}}} } 2}} \right.
 \kern-\nulldelimiterspace} 2}} \right){\rm{P}}^{\left( {{\mathop{\rm cov}} } \right)} \left( {{\rm{T}};G_{{\rm{BS}}}^{\left( {\max } \right)} G_{{\rm{MT}}}^{\left( {\min } \right)} } \right) \\
 & + \bar F_{\left| {\varepsilon _{{\rm{BS}}} } \right|} \left( {{{\omega _{{\rm{BS}}} } \mathord{\left/
 {\vphantom {{\omega _{{\rm{BS}}} } 2}} \right.
 \kern-\nulldelimiterspace} 2}} \right)F_{\left| {\varepsilon _{{\rm{MT}}} } \right|} \left( {{{\omega _{{\rm{MT}}} } \mathord{\left/
 {\vphantom {{\omega _{{\rm{MT}}} } 2}} \right.
 \kern-\nulldelimiterspace} 2}} \right){\rm{P}}^{\left( {{\mathop{\rm cov}} } \right)} \left( {{\rm{T}};G_{{\rm{BS}}}^{\left( {\min } \right)} G_{{\rm{MT}}}^{\left( {\max } \right)} } \right) \\ & + \bar F_{\left| {\varepsilon _{{\rm{BS}}} } \right|} \left( {{{\omega _{{\rm{BS}}} } \mathord{\left/
 {\vphantom {{\omega _{{\rm{BS}}} } 2}} \right.
 \kern-\nulldelimiterspace} 2}} \right)\bar F_{\left| {\varepsilon _{{\rm{MT}}} } \right|} \left( {{{\omega _{{\rm{MT}}} } \mathord{\left/
 {\vphantom {{\omega _{{\rm{MT}}} } 2}} \right.
 \kern-\nulldelimiterspace} 2}} \right){\rm{P}}^{\left( {{\mathop{\rm cov}} } \right)} \left( {{\rm{T}};G_{{\rm{BS}}}^{\left( {\min } \right)} G_{{\rm{MT}}}^{\left( {\min } \right)} } \right) \\
\end{split}
\end{equation}
\noindent where $\bar F_{\left| {\varepsilon _{\rm{q}} } \right|} \left( {{{\omega _q } \mathord{\left/ {\vphantom {{\omega _q } 2}} \right. \kern-\nulldelimiterspace} 2}} \right) = 1 - F_{\left| {\varepsilon _{\rm{q}} } \right|} \left( {{{\omega _q } \mathord{\left/ {\vphantom {{\omega _q } 2}} \right. \kern-\nulldelimiterspace} 2}} \right)$ for $q = \left\{ {{\rm{BS}},{\rm{MT}}} \right\}$. A similar expression holds for the rate.

\emph{Proof}: The proof immediately follows from Section \ref{PointingErrorModeling} and \eqref{Eq_3}. \hfill $\Box$
\end{proposition}
\begin{remark} \label{Remark_BeamErrors}
The coverage probability in \eqref{Eq_1__SecV} reduces to that computed in Section \ref{CoverageRate} in the absence of beamsteering errors. In this case, in fact, $\varepsilon _{{\rm{BS}}}  = \varepsilon _{{\rm{MT}}}  = 0$ and $F_{\left| {\varepsilon _q } \right|} \left( {{{\omega _q } \mathord{\left/ {\vphantom {{\omega _q } 2}} \right. \kern-\nulldelimiterspace} 2}} \right) = 1$ for $q = \left\{ {{\rm{BS}},{\rm{MT}}} \right\}$. Thus, as expected, ${\rm{P}}^{\left( {{\mathop{\rm cov}} } \right)} \left( {\rm{T}} \right) = {\rm{P}}^{\left( {{\mathop{\rm cov}} } \right)} \left( {{\rm{T}};G_{{\rm{BS}}}^{\left( {\max } \right)} G_{{\rm{MT}}}^{\left( {\max } \right)} } \right)$. \hfill $\Box$
\end{remark}
\subsection{Multi-Tier Cellular Deployment} \label{MultiTierCellular}
Consider a multi-tier mmWave cellular network, which is made of $\chi$ tiers of BSs. The BSs of each tier are distributed according to a homogeneous PPP of density $\lambda _k$ for $k=1, 2, \ldots, \chi$. The PPP of the $k$th tier is denoted by ${\Psi _{k} }$. Each tier of BSs is characterized by a different transmit power $\mathsf{P}_k$ and by different maximum and minimum directivity gains $G_{{\rm{BS}},k}^{\left( {\max } \right)}$ and $G_{{\rm{BS}},k}^{\left( {\min } \right)}$, respectively, for $k=1, 2, \ldots, \chi$. Beamsteering errors are not considered, since the generalization immediately follows from Section \ref{BeamformingErrors}. The MT is served by the BS providing the highest received power to it, by taking the transmit power and the directivity gain of the BSs into account. The BSs of each tier use the same carrier frequency (full-frequency reuse). Accordingly, path-loss, link state and fading models are the same for all tiers. In mathematical terms, and similar to Section \ref{CellAssociationReceivedPower}, the received SNR, under a noise-limited approximation, can be formulated as ${\rm{SNR = }}{{G_{{\rm{MT}}}^{\left( {\max } \right)} } \mathord{\left/ {\vphantom {{G_{{\rm{MT}}}^{\left( {\max } \right)} } {\left( {\sigma _N^2 P^{\left( 0 \right)} } \right)}}} \right. \kern-\nulldelimiterspace} {\left( {\sigma _N^2 P^{\left( 0 \right)} } \right)}}$, where:
\begin{equation}
\label{Eq_2__SecV}
\begin{split}
 & P^{\left( 0 \right)}  = \min \left\{ {\bigcup\limits_{k = 1}^\chi  {P_{k,{\rm{LOS}}}^{\left( 0 \right)} } ,\bigcup\limits_{k = 1}^\chi  {P_{k,{\rm{NLOS}}}^{\left( 0 \right)} } ,\bigcup\limits_{k = 1}^\chi  {P_{k,{\rm{OUT}}}^{\left( 0 \right)} } } \right\} \\
 & P_{k,s}^{\left( 0 \right)}  = \begin{cases}
 \mathop {\min }\limits_{n \in \Psi _{k,s} } \left\{ {\frac{{l_s \left( {r^{\left( {k,n} \right)} } \right)}}{{\left| {h_s^{\left( {k,n} \right)} } \right|^2 {\mathsf{P}}_k G_{{\rm{BS}},k}^{\left( {\max } \right)} }}} \right\} & \quad {\rm{if}}\quad \Psi _{k,s}  \ne \emptyset  \\
  + \infty & \quad {\rm{if}}\quad \Psi _{k,s}  = \emptyset  \\
 \end{cases} \\
\end{split}
\end{equation}
\noindent and $s = \left\{ {{\mathop{\rm LOS}\nolimits} ,{\rm{NLOS}}} \right\}$, $P_{k,{\rm{OUT}}}^{\left( 0 \right)}  =  + \infty$ for $k=1, 2, \ldots, \chi$, and $\Psi _{k,s}$ denotes the PPP of the BSs of tier $k$ that are in state $s$.
\begin{proposition} \label{MultiTier_Prop}
Let ${\rm{SNR = }}{{G_{{\rm{MT}}}^{\left( {\max } \right)} } \mathord{\left/ {\vphantom {{G_{{\rm{MT}}}^{\left( {\max } \right)} } {\left( {\sigma _N^2 P^{\left( 0 \right)} } \right)}}} \right. \kern-\nulldelimiterspace} {\left( {\sigma _N^2 P^{\left( 0 \right)} } \right)}}$, where $P^{\left( 0 \right)}$ is defined in \eqref{Eq_2__SecV}. Assume no beamsteering errors. The coverage probability in \eqref{Eq_8} can be formulated as follows:
\begin{equation}
\label{Eq_3__SecV}
\begin{split}
 {\rm{P}}^{\left( {{\mathop{\rm cov}} } \right)} \left( {\rm{T}} \right) & = F_{P^{\left( 0 \right)} } \left( {{{G_{{\rm{MT}}}^{\left( {\max } \right)} } \mathord{\left/
 {\vphantom {{G_{{\rm{MT}}}^{\left( {\max } \right)} } {\left( {\sigma _N^2 {\rm{T}}} \right)}}} \right.
 \kern-\nulldelimiterspace} {\left( {\sigma _N^2 {\rm{T}}} \right)}}} \right) = 1 - \exp \left( {\Lambda _P \left( {\left[ {0,{{G_{{\rm{MT}}}^{\left( {\max } \right)} } \mathord{\left/
 {\vphantom {{G_{{\rm{MT}}}^{\left( {\max } \right)} } {\left( {\sigma _N^2 {\rm{T}}} \right)}}} \right.
 \kern-\nulldelimiterspace} {\left( {\sigma _N^2 {\rm{T}}} \right)}}} \right)} \right)} \right) \\
 & \approx 1 - \exp \left( {\Lambda _P^{\left( {{\rm{approx}}} \right)} \left( {\left[ {0,{{G_{{\rm{MT}}}^{\left( {\max } \right)} } \mathord{\left/
 {\vphantom {{G_{{\rm{MT}}}^{\left( {\max } \right)} } {\left( {\sigma _N^2 {\rm{T}}} \right)}}} \right.
 \kern-\nulldelimiterspace} {\left( {\sigma _N^2 {\rm{T}}} \right)}}} \right)} \right)} \right) \\
\end{split}
\end{equation}
\noindent where $\Lambda _P \left(  \cdot  \right)$ and $\Lambda _P^{\left( {{\rm{approx}}} \right)} \left(  \cdot  \right)$ are defined as follows:
\begin{equation}
\label{Eq_4__SecV}
\begin{split}
 & \Lambda _P \left( {\left[ {0,x} \right)} \right) = \sum\limits_{k = 1}^\chi  {\sum\limits_{s = \left\{ {{\mathop{\rm LOS}\nolimits} ,{\rm{NLOS}}} \right\}} {{\mathbb{E}}_{\left| {h_s^{\left( 0 \right)} } \right|^2 } \left\{ {\Lambda _s \left( {\left[ {0,{\mathsf{P}}_k G_{{\rm{BS}},k}^{\left( {\max } \right)} \left| {h_s^{\left( 0 \right)} } \right|^2 x} \right)} \right)} \right\}} }  \\
 & \Lambda _P^{\left( {{\rm{approx}}} \right)} \left( {\left[ {0,x} \right)} \right) = \sum\limits_{k = 1}^\chi  {\sum\limits_{s = \left\{ {{\mathop{\rm LOS}\nolimits} ,{\rm{NLOS}}} \right\}} {\bar \Lambda _s^{\left( {{\rm{approx}}} \right)} \left( {\left[ {0,{\mathsf{P}}_k G_{{\rm{BS}},k}^{\left( {\max } \right)} x} \right)} \right)} }  \\
\end{split}
\end{equation}
\noindent and $\Lambda _s \left(  \cdot  \right)$ and $\bar \Lambda _s^{\left( {{\rm{approx}}} \right)} \left(  \cdot  \right)$ are defined in \eqref{Eq_11bis} and \eqref{Eq_2Ball__3}, respectively.

\emph{Proof}: The proof follows by using the same line of though as that of \textit{Lemma \ref{Intensity_Lemma}}. Since the $\chi$ PPPs are independent, the intensity of ${\bigcup\nolimits_{k = 1}^\chi  {P_{k,{\bar s}}^{\left( 0 \right)} } }$ for $\bar s = \left\{ {{\mathop{\rm LOS}\nolimits} ,{\rm{NLOS}},{\rm{OUT}}} \right\}$ is the summation of the intensities of the $\chi$ tiers. The intensity of each tier can be computed as in the proof of \textit{Lemma \ref{Intensity_Lemma}}, by taking into account that ${\mathsf{P}}_k$ and $G_{{\rm{BS}},k}^{\left( {\max } \right)}$ act as constants and ${\left| {h_s^{\left( {k,n} \right)} } \right|^2 }$ acts a as random variable for each tier. The proof is concluded by invoking \textit{Lemma \ref{CDF_Lemma}}. \hfill $\Box$
\end{proposition}

In summary, by capitalizing on the two-ball approximation introduced in Section \ref{TwoBall_Approximation}, an approximated closed-form expression of the coverage of general multi-tier mmWave cellular networks is provided. With the aid of \textit{Proposition \ref{BeamError_Prop}}, beamsteering errors can be taken into account, by still having a closed-form expression. The rate follows from \eqref{Eq_9} and, in general, a single integral needs to be computed. The closed-form mathematical formulation in \textit{Proposition \ref{MultiTier_Prop}} is based on two main assumptions: 1) mmWave systems are noise-limited and 2) empirically derived link state models are approximated by a two-ball link state model. The accuracy of these two approximations is investigated in the next section with the aid of Monte Carlo simulations.
\section{Numerical and Simulation Results} \label{Results}
In this section, we illustrate some numerical examples for validating the accuracy of the proposed mathematical frameworks and for comparing mmWave and $\mu$Wave cellular networks. The frameworks are substantiated with the aid of Monte Carlo simulations, where some modeling assumptions used for analytical tractability are not enforced in the system simulator. Notably, coverage and rate are computed under the noise-limited assumption in Section \ref{CoverageRate}. This approximation is \textit{not} retained in the system simulator, in order to show to which extent the noise-limited assumption holds for mmWave systems. Monte Carlo simulation results are obtained by using the system simulator described in \cite{MDR_TCOMrate}-\cite{MDR_COMMLPeng}, to which the reader is referred for further information.

Unless otherwise stated, the following setup is considered for obtaining the numerical examples, which agrees with previous studies in this field \cite{Rappaport__mmWaveJSAC}, \cite{Heath__mmWave}, \cite{Singh__mmWave}. In particular, channel and blockage models are taken from \cite{Rappaport__mmWaveJSAC}. In addition:
\begin{itemize}
  \item Two mmWave cellular networks are studied, which operate at a carrier frequency, $F_c$, equal to $F_c=28$ GHz and $F_c=73$ GHz. The transmission bandwidth is ${\rm{BW}} = 2$ GHz. The noise figure is ${\mathcal{N}}_{\rm{dB}}  = 10$. The transmit power for single-tier networks is $P=30$ dBm. The setup for multi-tier networks is summarized in the caption of the figures.
  \item The path-loss model is as follows \cite[Table I]{Rappaport__mmWaveJSAC}: $\alpha _{{\rm{LOS}}}  = 61.4$ dB, $\beta _{{\rm{LOS}}}  = 2$ and $\alpha _{{\rm{NLOS}}}  = 72$ dB, $\beta _{{\rm{NLOS}}}  = 2.92$ if $F_c=28$ GHz and $\alpha _{{\rm{LOS}}}  = 69.8$ dB, $\beta _{{\rm{LOS}}}  = 2$ and $\alpha _{{\rm{NLOS}}}  = 82.7$ dB, $\beta _{{\rm{NLOS}}}  = 2.69$ if $F_c=73$ GHz.
  \item The shadowing model is as follows \cite[Table I]{Rappaport__mmWaveJSAC}: $\sigma _{{\rm{LOS}}}^{(\rm{dB})}  = 5.8$, $\sigma _{{\rm{NLOS}}}^{(\rm{dB})}  = 8.7$ if $F_c=28$ GHz and $\sigma _{{\rm{LOS}}}^{(\rm{dB})}  = 5.8$, $\sigma _{{\rm{NLOS}}}^{(\rm{dB})}  = 8.7$ if $F_c=73$ GHz. On the other hand, $\mu^{(\rm{dB})}$ is assumed to be equal to zero for both LOS and NLOS scenarios.
  \item The blockage model is as follows \cite[Table I]{Rappaport__mmWaveJSAC}: $\delta _{{\rm{LOS}}}  = {1 \mathord{\left/ {\vphantom {1 {67.1}}} \right. \kern-\nulldelimiterspace} {67.1}}$, $\gamma _{{\rm{LOS}}}  = 1$ and $\delta _{{\rm{OUT}}}  = 5.2$, $\gamma _{{\rm{OUT}}}  = \exp \left( {{1 \mathord{\left/ {\vphantom {1 {30}}} \right. \kern-\nulldelimiterspace} {30}}} \right)$, for both $F_c=28$ GHz and $F_c=73$ GHz scenarios.
  \item The directional beamforming model for single-tier networks is as follows \cite{Heath__mmWave}: $G_{{\rm{BS}}}^{\left( {\max } \right)}  = G_{{\rm{MT}}}^{\left( {\max } \right)}  = 20$ dB, $G_{{\rm{BS}}}^{\left( {\min } \right)}  = G_{{\rm{MT}}}^{\left( {\min } \right)}  = -10$ dB and $\omega _{{\rm{BS}}}  = \omega _{{\rm{MT}}}  = 30$ degrees. The setup for multi-tier networks is summarized in the caption of the figures.
  \item Similar to \cite{Heath__mmWave}, the density of BSs, $\lambda$, is represented as a function of the average cell radius, \textit{i.e.}, $R_c  = \sqrt {{1 \mathord{\left/{\vphantom {1 {\left( {\pi \lambda } \right)}}} \right. \kern-\nulldelimiterspace} {\left( {\pi \lambda } \right)}}}$.
  \item As for $\mu$Wave cellular networks, a setup similar to \cite{Rappaport__mmWaveJSAC} is considered. In particular, we set $F_c = 2.5$ GHz, ${\rm{BW}} = 40$ MHz, $G_{{\rm{MT}}}^{\left( {\max } \right)}  = G_{{\rm{MT}}}^{\left( {\min } \right)}  = 0$ dB and $\omega _{{\rm{MT}}}  = 360$ degrees. The channel model is chosen as in \cite[Eq. (11)]{Rappaport__mmWaveJSAC}, \textit{i.e.}, $l\left( r \right)^{\left( {{\rm{dB}}} \right)}  = 22.7 + 36.7\log _{10} \left( r \right) + 26\log _{10} \left( {2.5} \right)$. All channels are assumed to be in a NLOS state, with a shadowing standard deviation equal to $\sigma _{{\rm{NLOS}}}  = 4$. No outage state is considered, \textit{i.e.}, $p_{{\rm{OUT}}} \left( r \right) = 0$. The rest of the paraments is the same as for the mmWave cellular network setup.
  \item As for the results obtained with the mathematical frameworks, the following holds. The curves related to the cell association based on the smallest path-loss are obtained by using the formulas in \textit{Proposition \ref{Pcov_Prop}} for the coverage and \eqref{Eq_9} for the rate. The curves related to the cell association based on the highest received power are obtained by using the formulas in \textit{Proposition \ref{PcovBest_Prop}} for the coverage and \eqref{Eq_9} for the rate. In this second case, only the formulas obtained by using the two-ball approximation are shown. As for the setups in Section \ref{Generalizations}, the formulas in \textit{Proposition \ref{BeamError_Prop}} and \textit{Proposition \ref{MultiTier_Prop}} are used. In all cases, the formulas obtained from the two-ball approximation are used.
\end{itemize}

\begin{figure}[!t]
\centering
\includegraphics [width=0.90\columnwidth] {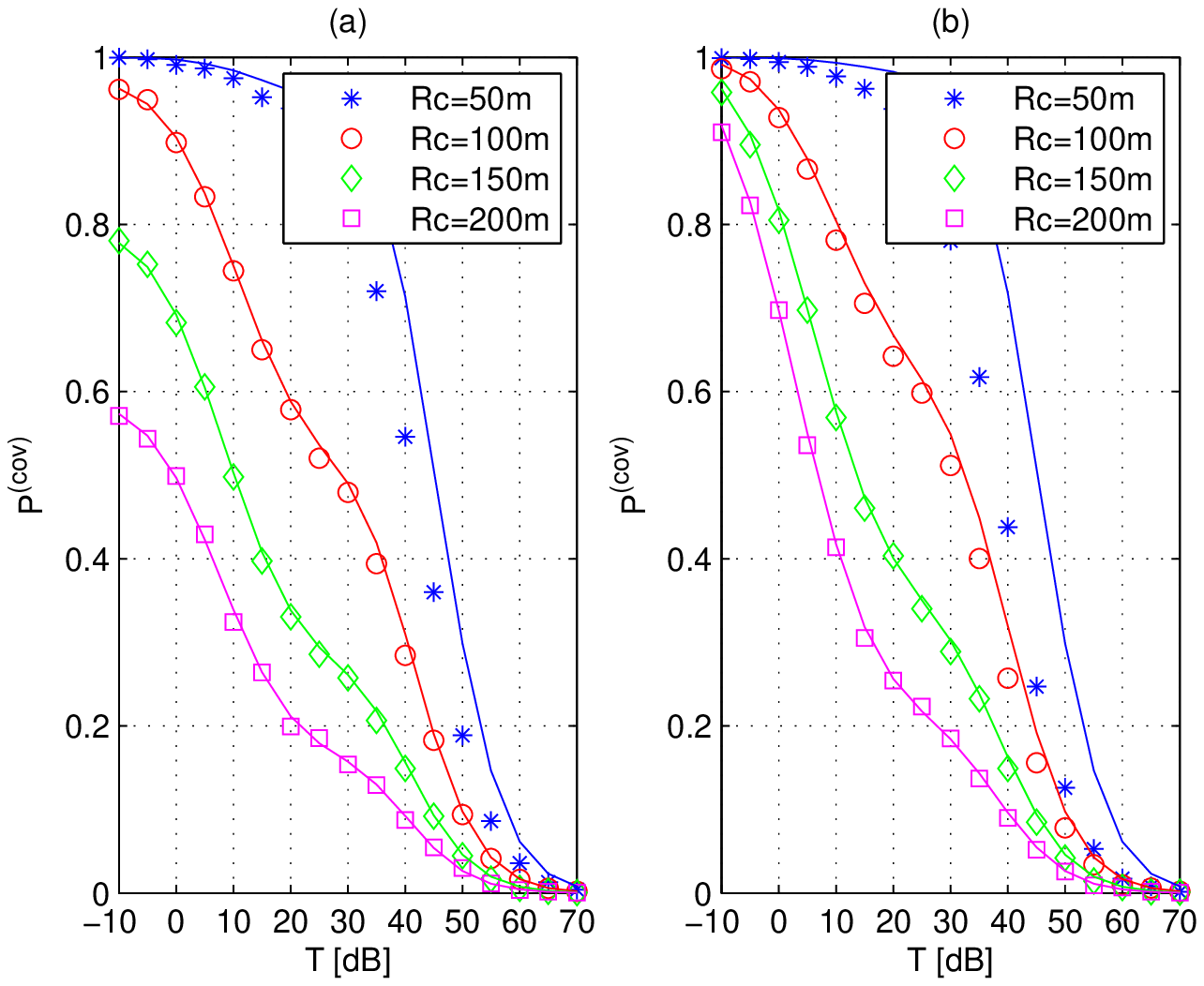}
\vspace{-0.40cm} \caption{\scriptsize{Coverage probability of a mmWave cellular network at $F_c=28$ GHz. Smallest path-loss cell association. (a) $p_{{\rm{OUT}}}(\cdot)$ in \eqref{Eq_4}. (b) $p_{{\rm{OUT}}}(r) = 0$. Solid lines: mathematical framework. Markers: Monte Carlo simulations.}} \label{Fig1_PL} \vspace{-0.40cm}
\end{figure}

\begin{figure}[!t]
\centering
\includegraphics [width=0.90\columnwidth] {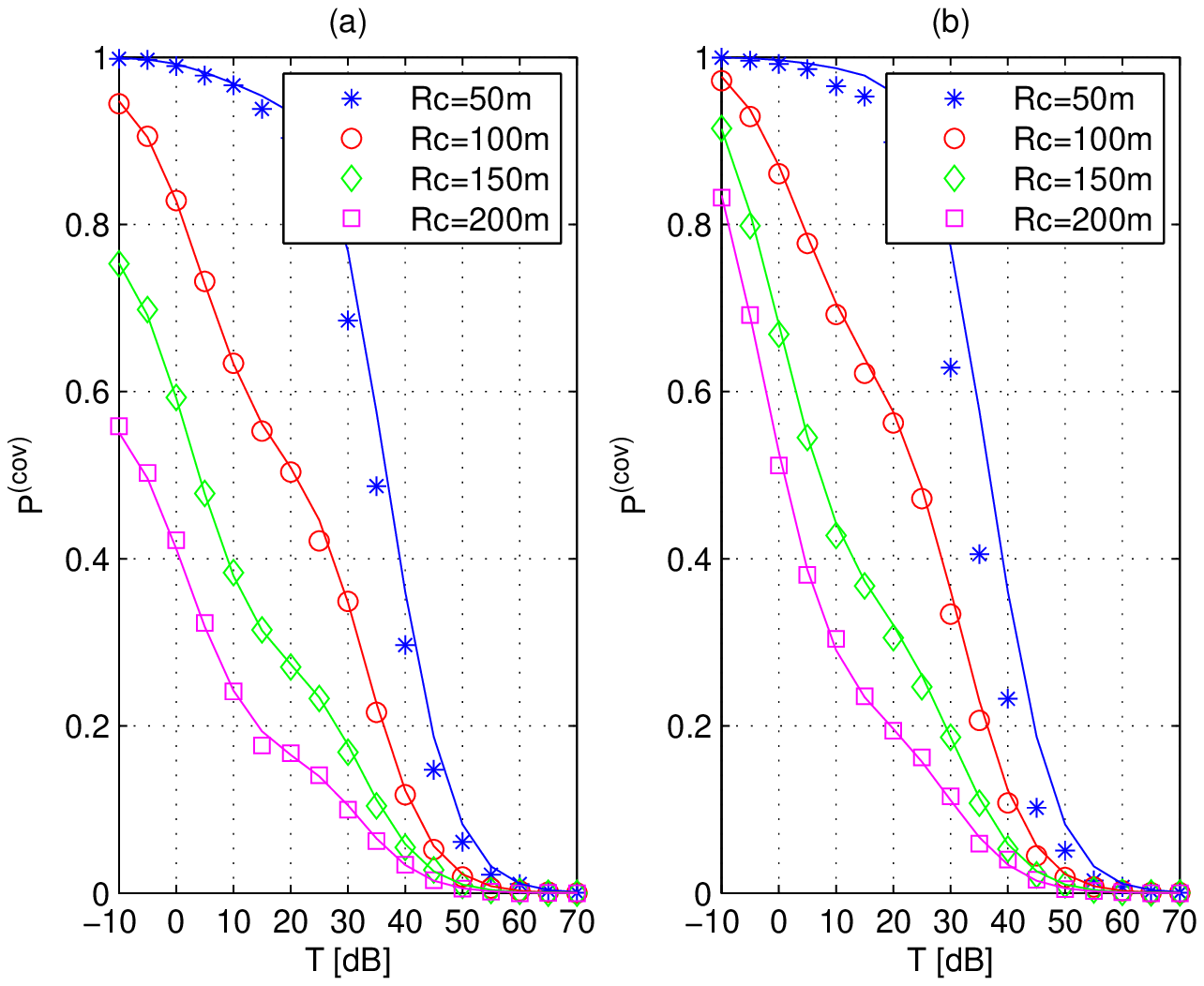}
\vspace{-0.40cm} \caption{\scriptsize{Coverage probability of a mmWave cellular network at $F_c=73$ GHz. Smallest path-loss cell association. (a) $p_{{\rm{OUT}}}(\cdot)$ in \eqref{Eq_4}. (b) $p_{{\rm{OUT}}}(r) = 0$. Solid lines: mathematical framework. Markers: Monte Carlo simulations.}} \label{Fig2_PL} \vspace{-0.40cm}
\end{figure}

\begin{figure}[!t]
\centering
\includegraphics [width=0.90\columnwidth] {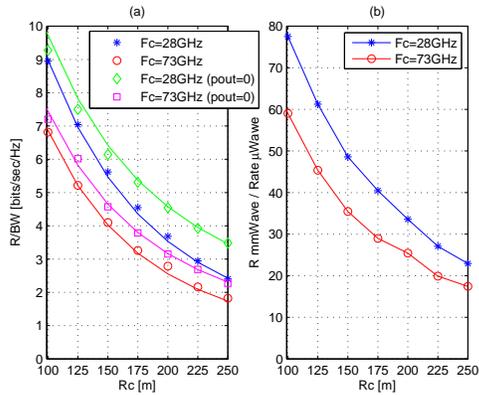}
\vspace{-0.40cm} \caption{\scriptsize{Average rate of a mmWave cellular network at $F_c=28$ GHz and $F_c=73$ GHz. Smallest path-loss cell association. (a) The normalized rate ${\rm{R}}/{\rm{BW}}$ is shown. Solid lines: mathematical framework. Markers: Monte Carlo simulations. (b) Ratio of the average rates of two mmWave networks at $F_c=28$ GHz and $F_c=73$ GHz and of a $\mu$Wave network at $F_c=2.5$ GHz.}} \label{Fig3_PL} \vspace{-0.40cm}
\end{figure}

\begin{figure}[!t]
\centering
\includegraphics [width=0.90\columnwidth] {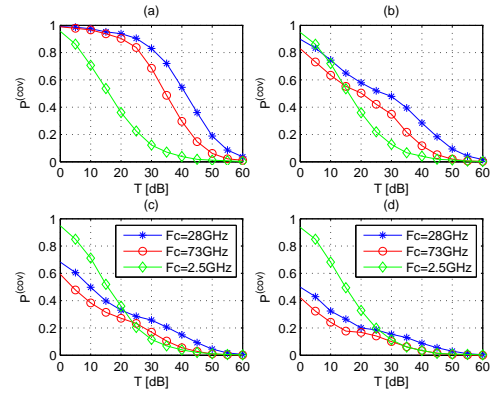}
\vspace{-0.40cm} \caption{\scriptsize{Coverage probability of mmWave and $\mu$Wave cellular networks at $F_c=28$ GHz (mmWave), $F_c=73$ GHz (mmWave) and $F_c=2.5$ GHz ($\mu$Wave). Smallest path-loss cell association. For mmWave networks, $p_{{\rm{OUT}}}(\cdot)$ in \eqref{Eq_4}. For the $\mu$Wave cellular network, $p_{{\rm{OUT}}}(r) = 0$. (a) $R_c = 50$ m. (b) $R_c = 100$ m. (c) $R_c = 150$ m. (d) $R_c = 200$ m.}} \label{Fig4_PL} \vspace{-0.40cm}
\end{figure}

\begin{figure}[!t]
\centering
\includegraphics [width=0.90\columnwidth] {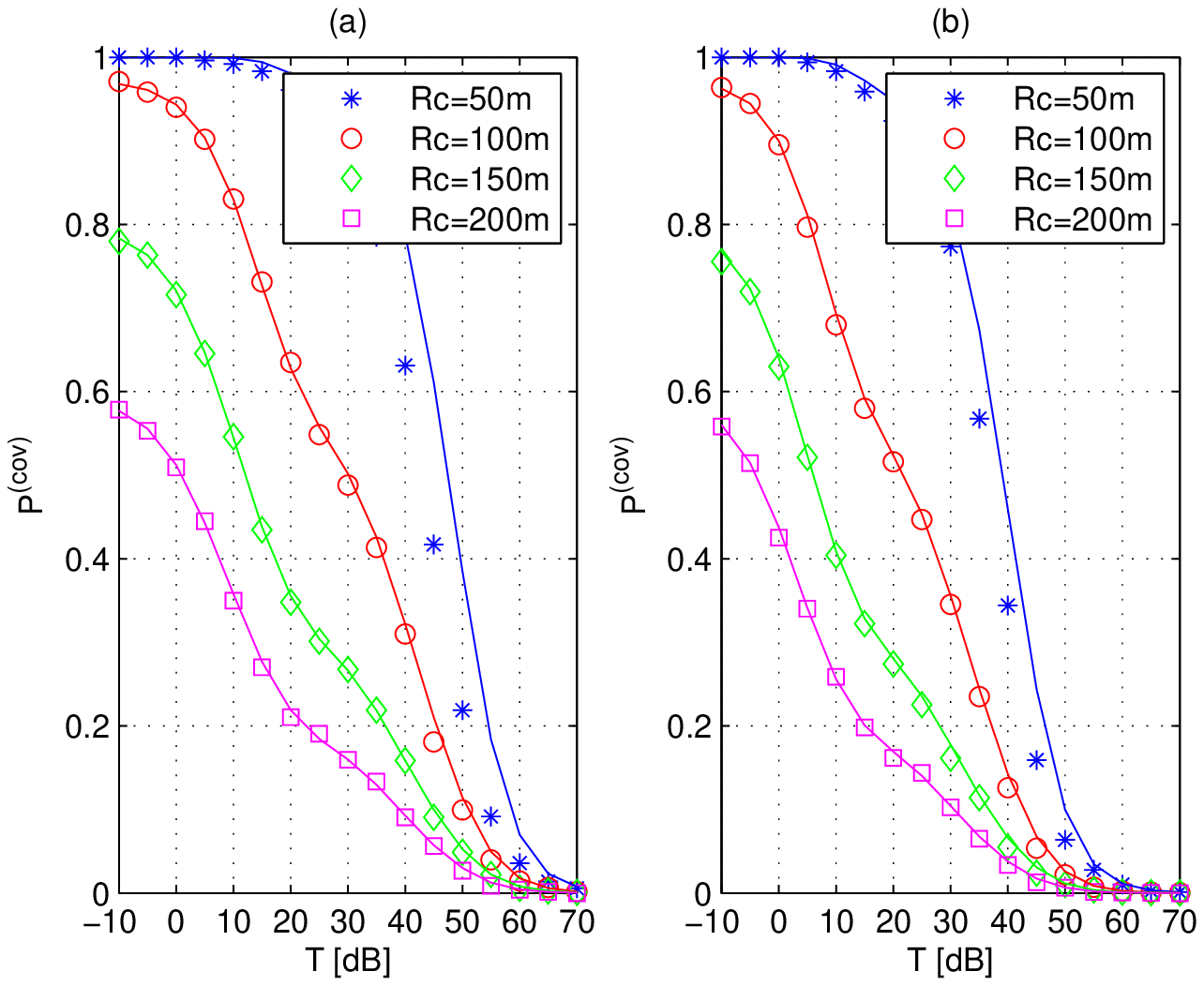}
\vspace{-0.40cm} \caption{\scriptsize{Coverage probability of a mmWave cellular network at $F_c=28$ GHz (a) and $F_c=73$ GHz (b). Highest received power cell association. $p_{{\rm{OUT}}}(\cdot)$ in \eqref{Eq_4}. Solid lines: mathematical framework. Markers: Monte Carlo simulations.}} \label{Fig1_Best} \vspace{-0.40cm}
\end{figure}

\begin{figure}[!t]
\centering
\includegraphics [width=0.90\columnwidth] {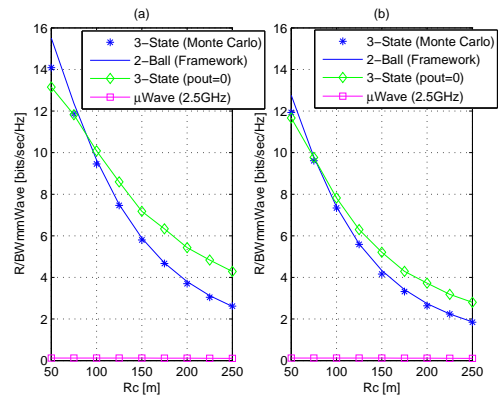}
\vspace{-0.40cm} \caption{\scriptsize{Average rate of a mmWave cellular network at $F_c=28$ GHz (a) and $F_c=73$ GHz (b). Highest received power cell association. $p_{{\rm{OUT}}}(\cdot)$ in \eqref{Eq_4}. The normalized rate ${\rm{R}}/{\rm{BW}}$ is shown. Solid lines: mathematical framework. Markers: Monte Carlo simulations. The figure also shows the average rate of the mmWave cellular networks without outage state ($p_{{\rm{OUT}}}(\cdot) = 0$) and that of a $\mu$Wave cellular network at $F_c=2.5$ GHz. All rates are normalized to the transmission bandwidth of mmWave cellular networks.}} \label{Fig2_Best} \vspace{-0.40cm}
\end{figure}

\begin{figure}[!t]
\centering
\includegraphics [width=0.90\columnwidth] {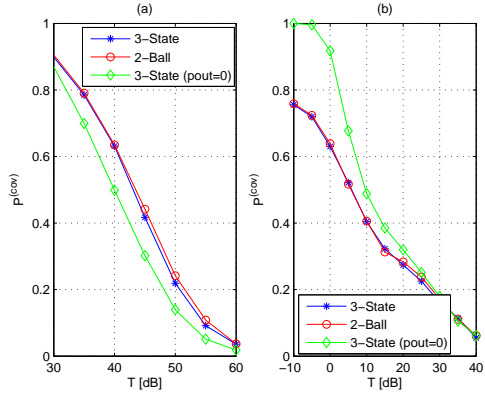}
\vspace{-0.40cm} \caption{\scriptsize{Coverage probability of a mmWave cellular network: Impact of the outage state. Highest received power cell association. $p_{{\rm{OUT}}}(\cdot)$ in \eqref{Eq_4}. (a) $F_c=28$ GHz and $R_c = 50$ m. (b) $F_c=73$ GHz and $R_c = 150$ m.}} \label{Fig3_Best} \vspace{-0.40cm}
\end{figure}

\begin{figure}[!t]
\centering
\includegraphics [width=0.90\columnwidth] {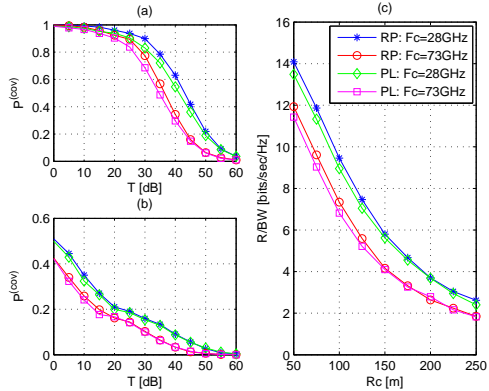}
\vspace{-0.40cm} \caption{\scriptsize{Coverage probability (a, b) and average rate (c) of a mmWave cellular network: Impact of cell association. PL means based on the smallest path-loss and RP means based on the highest received power. $p_{{\rm{OUT}}}(\cdot)$ in \eqref{Eq_4}. (a) $R_c = 50$ m. (b) $R_c = 200$ m. (c) The normalized rate ${\rm{R}}/{\rm{BW}}$ is shown.}} \label{Fig4_Best} \vspace{-0.40cm}
\end{figure}

\begin{figure}[!t]
\centering
\includegraphics [width=0.90\columnwidth] {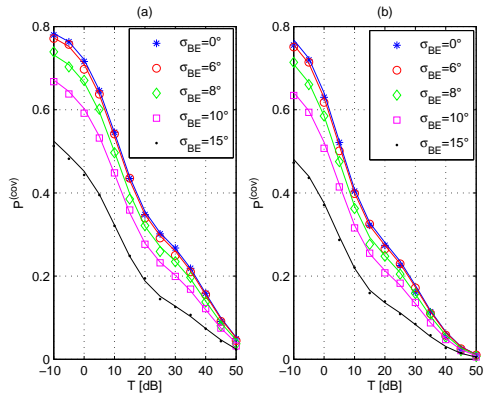}
\vspace{-0.40cm} \caption{\scriptsize{Coverage probability of a mmWave cellular network with $R_c = 150$ m at $F_c=28$ GHz (a) and $F_c=73$ GHz (b): Impact of beamsteering errors. Highest received power cell association. $p_{{\rm{OUT}}}(\cdot)$ in \eqref{Eq_4}. Solid lines: mathematical framework. Markers: Monte Carlo simulations.}} \label{Fig1_BeamError} \vspace{-0.40cm}
\end{figure}

\begin{figure}[!t]
\centering
\includegraphics [width=0.90\columnwidth] {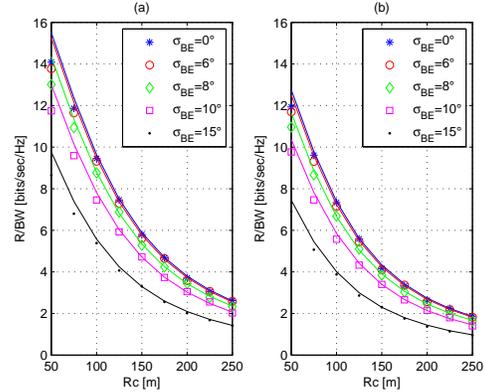}
\vspace{-0.43cm} \caption{\scriptsize{Average rate of a mmWave cellular network at $F_c=28$ GHz (a) and $F_c=73$ GHz (b): Impact of beamsteering errors. Highest received power cell association. $p_{{\rm{OUT}}}(\cdot)$ in \eqref{Eq_4}. The normalized rate ${\rm{R}}/{\rm{BW}}$ is shown. Solid lines: mathematical framework. Markers: Monte Carlo simulations.}} \label{Fig2_BeamError} \vspace{-0.55cm}
\end{figure}

\begin{figure}[!t]
\centering
\includegraphics [width=0.90\columnwidth] {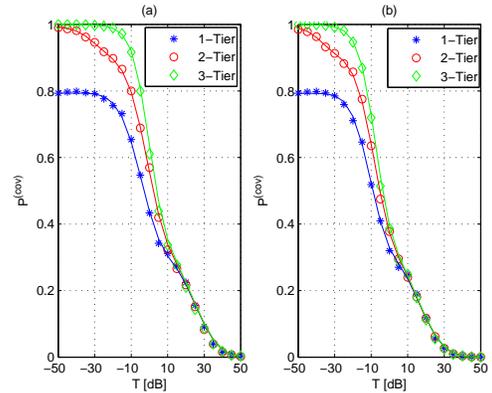}
\vspace{-0.43cm} \caption{\scriptsize{Coverage probability of a mmWave cellular network at $F_c=28$ GHz (a) and $F_c=73$ GHz (b). Cell association in \eqref{Eq_2__SecV}. $p_{{\rm{OUT}}}(\cdot)$ in \eqref{Eq_4}. Solid lines: mathematical framework. Markers: Monte Carlo simulations. Setup. Tier-1: $R_c = 150$ m, $\mathsf{P}=30$ dBm, $G_{{\rm{BS}}}^{\left( {\max } \right)}  = 20$ dB, $G_{{\rm{BS}}}^{\left( {\min } \right)}= -10$ dB, $\omega_{\rm{BS}} = 30^{\circ}$. Tier-2: $R_c = 100$ m, $\mathsf{P}=10$ dBm, $G_{{\rm{BS}}}^{\left( {\max } \right)} = 10$ dB, $G_{{\rm{BS}}}^{\left( {\min } \right)} = 0$ dB, $\omega_{\rm{BS}} = 40^{\circ}$. Tier-3: $R_c = 50$ m, $\mathsf{P}=5$ dBm, $G_{{\rm{BS}}}^{\left( {\max } \right)} = 5$ dB, $G_{{\rm{BS}}}^{\left( {\min } \right)} = 0$ dB, $\omega_{\rm{BS}} = 50^{\circ}$. Also, $G_{{\rm{MT}}}^{\left( {\max } \right)}  = 5$ dB, $G_{{\rm{MT}}}^{\left( {\min } \right)}  = 0$ dB, $\omega_{\rm{MT}} = 50^{\circ}$.}} \label{Fig1_MultiTier} \vspace{-0.55cm}
\end{figure}

\begin{figure}[!t]
\centering
\includegraphics [width=0.90\columnwidth] {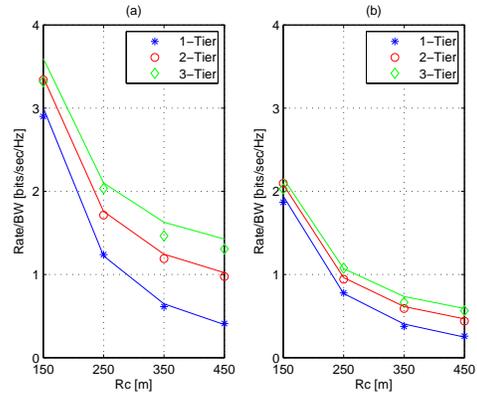}
\vspace{-0.43cm} \caption{\scriptsize{Average rate of a mmWave cellular network at $F_c=28$ GHz (a) and $F_c=73$ GHz (b). Cell association in \eqref{Eq_2__SecV}. $p_{{\rm{OUT}}}(\cdot)$ in \eqref{Eq_4}. Solid lines: mathematical framework. Markers: Monte Carlo simulations. The same setup as in Fig. \ref{Fig1_MultiTier} is considered with an exception. The values of $R_c$ shown in the figure are related to Tier-1. The cell radii of Tier-2 and Tier-3 are kept fixed to 100 and 50 meters.}} \label{Fig2_MultiTier} \vspace{-0.40cm}
\end{figure}

Selected numerical results are illustrated in Figs. \ref{Fig1_PL}-\ref{Fig2_MultiTier}. From these figures, we observe that the proposed noise-limited approximation is quite accurate for practical densities of BSs. If $R_c \ge 100$ meters for the considered setup, in particular, we observe that mmWave cellular networks can be assumed to be noise-limited. If the density of BSs increases, on the other hand, this approximation may no longer hold. The performance gap compared to Monte Carlo simulations is, however, tolerable and this shows that, in any case, mmWave cellular networks are likely not to be interference-limited. This finding is in agreement with recent published papers that considered a simplified blockage model \cite{Singh__mmWave}. The figures also show that, in general, the presence of an outage state reduces the coverage probability. This is noticeable, in particular, for small values of the reliability threshold $\rm{T}$. Furthermore, as expected, the performance gets better as the average cell radius $R_c$ decreases, \textit{i.e.}, for denser network deployments. Cell associations based on the smallest path-loss and the highest received power provide, in general, very close performance. Some figures deserve some additional comments.

In Fig. \ref{Fig4_PL}, mmWave and $\mu$Wave cellular networks are compared by assuming a cell association based on the smallest path-loss. This figure shows that mmWave systems have the potential of outperforming $\mu$Wave systems, provided that the network density is sufficiently high. Otherwise, $\mu$Wave systems are still to be preferred, especially for small values of the reliability threshold $\rm{T}$. As expected, mmWave transmission at $F_c=28$ GHz slightly outperforms its counterpart at $F_c=73$ GHz due to a smaller path-loss.

In Figs. \ref{Fig3_PL} and \ref{Fig2_Best}, the rate of mmWave and $\mu$Wave networks is compared. They show that mmWave networks are capable of significantly enhancing the average rate. This is mainly due to the larger transmission bandwidth, which is 50 times larger, in the considered setup, for mmWave systems. The figure shows, however, that the gain can be larger than the ratio of the bandwidths, especially for medium/dense cellular deployments. Figure \ref{Fig2_Best} shows an interesting phenomenon: for dense network deployments, \textit{i.e.}, $R_c<100$ meters, the average rate may be larger in the presence of an outage state. This is because the BSs that are in outage do not contribute to the other-cell interference. On the other hand, the outage state negatively affects the rate if the BSs are sparely deployed. A similar trend emerges in Fig. \ref{Fig3_Best} for the coverage probability.

In Figs. \ref{Fig1_BeamError} and \ref{Fig2_BeamError}, the impact of beamsteering errors is investigated. The figures confirm that beamsteering errors degrade, in general, the achievable performance. In the considered setup, the degradation is noticeable if the standard deviation of the pointing error is greater than 6 degrees.

Finally, Figs. \ref{Fig1_MultiTier} and \ref{Fig2_MultiTier} confirm that multi-tier networks provide better performance, especially for small values of the reliability threshold $\rm{T}$ and for large cell radii of the higher tier of BSs. In spite of the large number of deployed BSs in this setup and the small cell radius for the lowest tier of BSs ($R_c = 50$ meters for Tier-3), the results confirm that the noise-limited approximation still holds for mmWave cellular networks. This occurs even though the directivity gain of the lower tiers of BSs and of the MT is not that high.
\section{Conclusion} \label{Conclusion}
In the present paper, a new analytical framework for computing coverage probability and average rate of mmWave cellular networks has been proposed. Its novelty lies in taking into account realistic channel and blockage models for mmWave propagation, which are based on empirical data available in the literature. A systematic two-ball approximation for modeling the link-state of mmWave communications is introduced, which is based on matching the intensities of the PPPs of empirical three-state and approximated two-ball link models. The proposed mathematical methodology relies on the noise-limited assumption for modeling mmWave cellular systems, which is shown to be sufficiently accurate for typical densities of BSs and for envisioned transmission bandwidths. The proposed approach is applicable to different cell association criteria, to multi-tier cellular deployments and it accounts for beamforming pointing errors. The numerical examples have confirmed that sufficiently dense mmWave cellular networks have the inherent capability of outperforming their $\mu$Wave counterpart.
%
%
%
%
%
\section*{Appendix I -- Proofs of the Results in Section \ref{PPP_PathLoss}} \label{Appendix_I}
\subsection{Proof of Lemma \ref{Intensity_Lemma}}
The proof follows by using a methodology similar to \cite[Sec. II-A]{Blaszczyszyn_Infocom2013}. In particular, by invoking the displacement theorem of PPPs \cite[Th. 1.10]{BaccelliBook2009}, the process of the scaled propagation losses $\Phi  = \left\{ {{{l\left( {r^{\left( n \right)} } \right)} \mathord{\left/ {\vphantom {{l\left( {r^{\left( n \right)} } \right)} {{\mathcal{A}}^{\left( n \right)} }}} \right. \kern-\nulldelimiterspace} {{\mathcal{A}}^{\left( n \right)} }},n \in \Psi } \right\}$ can be interpreted as a transformation of $\Psi$, which is still a PPP on $\mathbb{R}^+$. From Section \ref{LinkStateModeling}, we know that $\Psi  = \Psi _{{\rm{LOS}}}  \cup \Psi _{{\rm{NLOS}}}  \cup \Psi _{{\rm{OUT}}}$. Since $\Psi _{{\rm{LOS}}}$, $\Psi _{{\rm{NLOS}}}$ and $\Psi _{{\rm{OUT}}}$ are independent, the density (or intensity), ${\Lambda_{\Phi} \left( \cdot \right)}$, of $\Phi  = \left\{ {{{l\left( {r^{\left( n \right)} } \right)} \mathord{\left/ {\vphantom {{l\left( {r^{\left( n \right)} } \right)} {{\mathcal{A}}^{\left( n \right)} }}} \right. \kern-\nulldelimiterspace} {{\mathcal{A}}^{\left( n \right)} }},n \in \Psi } \right\}$ is equal to the summation of the intensities of $\Phi _{{\rm{LOS}}}$, $\Phi _{{\rm{NLOS}}}$ and $\Phi _{{\rm{OUT}}}$. Since the path-loss of the links in outage is infinite, by definition its intensity is equal to zero. The intensities, $\Lambda _{\Phi _{{\rm{LOS}}} } \left(  \cdot  \right)$ and $\Lambda _{\Phi _{{\rm{NLOS}}} } \left(  \cdot  \right)$ of $\Phi _{{\rm{LOS}}}$ and $\Phi _{{\rm{NLOS}}}$, respectively, on the other hand, can be computed by using mathematical steps similar to the proof of \cite[Lemma 1]{Blaszczyszyn_Infocom2013}. More specifically, we have:
\begin{equation}
\label{Eq_App1__1}
\Lambda _{\Phi _s } \left( {\left[ {0,x} \right)} \right) = 2\pi \lambda {\mathbb{E}}_{{\mathcal{A}}_s } \left\{ {\int\nolimits_0^{ + \infty } {{\mathcal{H}}\left( {{\mathcal{A}}_s x - \left( {\kappa _s r} \right)^{\beta _s } } \right)p_s \left( r \right)rdr} } \right\}
\end{equation}
\noindent where $p_s \left( \cdot \right)$ for $s = \left\{ {{\rm{LOS}},{\rm{NLOS}}} \right\}$ is defined in \eqref{Eq_4}.

Equation \eqref{Eq_11} follows by inserting $p_s \left(  \cdot  \right)$ of \eqref{Eq_4} in \eqref{Eq_App1__1} and by computing the integrals with the aid of the notable result $\int\nolimits_a^b {e^{ - cr} rdr}  = \left( {{1 \mathord{\left/ {\vphantom {1 {c^2 }}} \right.\kern-\nulldelimiterspace} {c^2 }}} \right)\left( {e^{ - ca}  + ae^{ - ca}  - e^{ - cb}  - be^{ - cb} } \right)$.
\subsection{Proof of Corollary \ref{Intensity2Ball_Corollary}}
It follows by calculating the expectation of \eqref{Eq_2Ball__2}, where $x$ is replaced with ${{\mathcal{A}}_s }x$, with respect to ${{\mathcal{A}}_s }$, by using the results:
\begin{equation}
\label{Eq_App1__2}
\begin{split}
 & F_{{\mathcal{A}}_s } \left( y \right) = \Pr \left\{ {{\mathcal{A}}_s  \le y} \right\} = \int\nolimits_0^y {f_{{\mathcal{A}}_s } \left( \xi  \right)d\xi }  = {1 \mathord{\left/
 {\vphantom {1 2}} \right.
 \kern-\nulldelimiterspace} 2} + {1 \mathord{\left/
 {\vphantom {1 2}} \right.
 \kern-\nulldelimiterspace} 2}{\rm{erf}}\left( {{{\left( {\ln \left( y \right) - \mu _s } \right)} \mathord{\left/
 {\vphantom {{\left( {\ln \left( y \right) - \mu _s } \right)} {\left( {\sqrt 2 \sigma _s } \right)}}} \right.
 \kern-\nulldelimiterspace} {\left( {\sqrt 2 \sigma _s } \right)}}} \right) \\
 & m_{{\mathcal{A}}_s } \left( {\nu ,y} \right) = \int\nolimits_0^y {\xi ^\nu  f_{{\mathcal{A}}_s } \left( \xi  \right)d\xi }  = \left( {{1 \mathord{\left/
 {\vphantom {1 2}} \right.
 \kern-\nulldelimiterspace} 2}} \right)\exp \left\{ {\nu \mu _s  + \left( {{1 \mathord{\left/
 {\vphantom {1 2}} \right.
 \kern-\nulldelimiterspace} 2}} \right)\nu ^2 \sigma _s^2 } \right\} \\ & \hspace{5.06cm} \times {\rm{erfc}}\left( {{{\nu \sigma _s } \mathord{\left/
 {\vphantom {{\nu \sigma _s } {\sqrt 2 }}} \right.
 \kern-\nulldelimiterspace} {\sqrt 2 }} - {{\left( {\ln \left( y \right) - \mu _s } \right)} \mathord{\left/
 {\vphantom {{\left( {\ln \left( y \right) - \mu _s } \right)} {\left( {\sqrt 2 \sigma _s } \right)}}} \right.
 \kern-\nulldelimiterspace} {\left( {\sqrt 2 \sigma _s } \right)}}} \right) \\
 & \bar m_{{\mathcal{A}}_s } \left( {\nu ,y} \right) = \int\nolimits_y^{ + \infty } {\xi ^\nu  f_{{\mathcal{A}}_s } \left( \xi  \right)d\xi }  = \exp \left\{ {\nu \mu _s  + \left( {{1 \mathord{\left/
 {\vphantom {1 2}} \right.
 \kern-\nulldelimiterspace} 2}} \right)\nu ^2 \sigma _s^2 } \right\} - m_{{\mathcal{A}}_s } \left( {\nu ,y} \right) \\
 \end{split}
\end{equation}
\noindent where $f_{{\mathcal{A}}_s } \left( \xi  \right) = \left( {\sqrt {2\pi } \sigma _s \xi } \right)^{ - 1} \exp \left( { - {{\left( {\ln \left( \xi  \right) - \mu _s } \right)^2 } \mathord{\left/ {\vphantom {{\left( {\ln \left( \xi  \right) - \mu _s } \right)^2 } {\left( {2\sigma _s^2 } \right)}}} \right. \kern-\nulldelimiterspace} {\left( {2\sigma _s^2 } \right)}}} \right)$.
\subsection{Proof of Lemma \ref{BlockageProbability_Lemma}}
Since ${\Psi _{{\rm{LOS}}} }$ and ${\Psi _{{\rm{NLOS}}} }$ are independent, then ${\mathcal{P}}_{{\rm{blockage}}}  = \Pr \left\{ {\Psi _{{\rm{LOS}}}  = \emptyset  \cap \Psi _{{\rm{NLOS}}}  = \emptyset } \right\} = \Pr \left\{ {\Psi _{{\rm{LOS}}}  = \emptyset } \right\}\Pr \left\{ {\Psi _{{\rm{NLOS}}}  = \emptyset } \right\}$. From the void probability theorem of PPPs \cite{BaccelliBook2009}, we have:
\begin{equation}
\label{Eq_App1__1bis}
\Pr \left\{ {\Psi _s  = \emptyset } \right\} = \exp \left( { - 2\pi \lambda \int\nolimits_0^{ + \infty } {p_s \left( r \right)rdr} } \right)
\end{equation}
\noindent for $s = \left\{ {{\rm{LOS}},{\rm{NLOS}}} \right\}$ and $p_s \left( \cdot \right)$ is defined in \eqref{Eq_4}. The integral in \eqref{Eq_App1__1bis} can be computed in closed-form from \eqref{Eq_App1__1} by letting $x \to  + \infty$ for $\mathcal{A}_s =1$. The proof follows with the aid of some simplifications. Alternatively, the proof may be obtained directly from \textit{Lemma \ref{Intensity_Lemma}}. By definition of communication blockage, in fact, the equalities ${\mathcal{P}}_{{\rm{blockage}}}  = \Pr \left\{ {\Phi^{\left( 0 \right)}  =  + \infty } \right\} = \Pr \left\{ {L^{\left( 0 \right)}  \ge  + \infty } \right\} = 1 - F_{L^{\left( 0 \right)} } \left( {x \to  + \infty } \right)$ hold, from which the proof follows setting $\mathcal{A}_s =1$.
\section*{Appendix II -- Proofs of the Results in Section \ref{CoverageRate}} \label{Appendix_II}
\subsection{Proof of Proposition \ref{Pcov_Prop}}
From \eqref{Eq_8} and \eqref{Eq_1__SecIV}, the coverage probability can be formulated, by definition, as follows:
\begin{equation}
\label{Eq_App2__1}
\begin{split}
 {\rm{P}}^{\left( {{\mathop{\rm cov}} } \right)} \left( {\rm{T}} \right) & = {\mathbb{E}}_{L_{{\rm{LOS}}}^{\left( 0 \right)} } \left\{ {\Pr \left\{ {\left. {\frac{{{\mathsf{P}}G^{\left( 0 \right)} \left| {h_{{\rm{LOS}}}^{\left( 0 \right)} } \right|^2 }}{{\sigma _N^2 L_{{\rm{LOS}}}^{\left( 0 \right)} }} > {\rm{T}}} \right|L_{{\rm{LOS}}}^{\left( 0 \right)} } \right\}\Pr \left\{ {\left. {L_{{\rm{NLOS}}}^{\left( 0 \right)}  > L_{{\rm{LOS}}}^{\left( 0 \right)} } \right|L_{{\rm{LOS}}}^{\left( 0 \right)} } \right\}} \right\} \\
 & + {\mathbb{E}}_{L_{{\rm{NLOS}}}^{\left( 0 \right)} } \left\{ {\Pr \left\{ {\left. {\frac{{{\mathsf{P}}G^{\left( 0 \right)} \left| {h_{{\rm{NLOS}}}^{\left( 0 \right)} } \right|^2 }}{{\sigma _N^2 L_{{\rm{NLOS}}}^{\left( 0 \right)} }} > {\rm{T}}} \right|L_{{\rm{NLOS}}}^{\left( 0 \right)} } \right\}\Pr \left\{ {\left. {L_{{\rm{LOS}}}^{\left( 0 \right)}  > L_{{\rm{NLOS}}}^{\left( 0 \right)} } \right|L_{{\rm{NLOS}}}^{\left( 0 \right)} } \right\}} \right\} \\
 \end{split}
\end{equation}

Denote the first and second addends in \eqref{Eq_App2__1} by ${\rm{P}}_s^{\left( {{\mathop{\rm cov}} } \right)} \left( \cdot \right)$, where $s = {\rm{LOS}}$ and $s = {\rm{NLOS}}$, respectively. It can be computed by using the following results:
\begin{equation}
\label{Eq_App2__2}
\begin{split}
 & \Pr \left\{ {\left. {\left| {h_{\rm{s}}^{\left( 0 \right)} } \right|^2  > {{L_{\rm{s}}^{\left( 0 \right)} {\rm{T}}} \mathord{\left/
 {\vphantom {{L_{\rm{s}}^{\left( 0 \right)} {\rm{T}}} {\gamma ^{\left( 0 \right)} }}} \right.
 \kern-\nulldelimiterspace} {\gamma ^{\left( 0 \right)} }}} \right|L_{\rm{s}}^{\left( 0 \right)} } \right\}\mathop  = \limits^{\left( a \right)} {1 \mathord{\left/
 {\vphantom {1 2}} \right.
 \kern-\nulldelimiterspace} 2} - ({1 \mathord{\left/
 {\vphantom {1 2}} \right.
 \kern-\nulldelimiterspace} 2}){\rm{erf}}\left( {{{\left( {\ln \left( {{{L_{\rm{s}}^{\left( 0 \right)} {\rm{T}}} \mathord{\left/
 {\vphantom {{L_{\rm{s}}^{\left( 0 \right)} {\rm{T}}} {\gamma ^{\left( 0 \right)} }}} \right.
 \kern-\nulldelimiterspace} {\gamma ^{\left( 0 \right)} }}} \right) - \mu _s } \right)} \mathord{\left/
 {\vphantom {{\left( {\ln \left( {{{L_{\rm{s}}^{\left( 0 \right)} {\rm{T}}} \mathord{\left/
 {\vphantom {{L_{\rm{s}}^{\left( 0 \right)} {\rm{T}}} {\gamma ^{\left( 0 \right)} }}} \right.
 \kern-\nulldelimiterspace} {\gamma ^{\left( 0 \right)} }}} \right) - \mu _s } \right)} {\left( {\sqrt 2 \sigma _s } \right)}}} \right.
 \kern-\nulldelimiterspace} {\left( {\sqrt 2 \sigma _s } \right)}}} \right) \\
 & \Pr \left\{ {\left. {L_{{\rm{NLOS}}}^{\left( 0 \right)}  > L_{{\rm{LOS}}}^{\left( 0 \right)} } \right|L_{{\rm{LOS}}}^{\left( 0 \right)} } \right\}\mathop  = \limits^{\left( b \right)} \exp \left( { - \Lambda _{{\rm{NLOS}}} \left( {\left[ {0,L_{{\rm{LOS}}}^{\left( 0 \right)} } \right)} \right)} \right) \\
 & \Pr \left\{ {\left. {L_{{\rm{LOS}}}^{\left( 0 \right)}  > L_{{\rm{NLOS}}}^{\left( 0 \right)} } \right|L_{{\rm{NLOS}}}^{\left( 0 \right)} } \right\}\mathop  = \limits^{\left( c \right)} \exp \left( { - \Lambda _{{\rm{LOS}}} \left( {\left[ {0,L_{{\rm{NLOS}}}^{\left( 0 \right)} } \right)} \right)} \right) \\
 \end{split}
\end{equation}
\noindent where (a) follows from \eqref{Eq_App1__2}, and (b) and (c) follow from \textit{Lemma \ref{Intensity_Lemma}}, \textit{Lemma \ref{CDF_Lemma}} and \textit{Remark \ref{Remark_GeneralPPP}}, since $L^{\left( 0 \right)}  = \left. {\Phi ^{\left( 0 \right)} } \right|_{{\mathcal{A}}_{{\rm{LOS}}}  = {\mathcal{A}}_{{\rm{NLOS}}}  = 1}  = \min \left\{ {L_{{\rm{LOS}}}^{\left( 0 \right)} ,L_{{\rm{NLOS}}}^{\left( 0 \right)} } \right\}$, $L_{{\rm{LOS}}}^{\left( 0 \right)}  = \left. {\Phi _{{\rm{LOS}}}^{\left( 0 \right)} } \right|_{{\mathcal{A}}_{{\rm{LOS}}}  = 1}$ and $L_{{\rm{NLOS}}}^{\left( 0 \right)}  = \left. {\Phi _{{\rm{NLOS}}}^{\left( 0 \right)} } \right|_{{\mathcal{A}}_{{\rm{NLOS}}}  = 1}$. The proof follows by explicitly writing the expectation with respect to $L_{{\rm{LOS}}}^{\left( 0 \right)}$ and $L_{{\rm{NLOS}}}^{\left( 0 \right)}$ in terms of their PDFs, which can be formulated, similar to (b) and (c), as $f_{L_s^{\left( 0 \right)} } \left( \xi  \right) = {{d\Pr \left\{ {L_s^{\left( 0 \right)}  < \xi } \right\}} \mathord{\left/ {\vphantom {{d\Pr \left\{ {L_s^{\left( 0 \right)}  < \xi } \right\}} {d\xi }}} \right. \kern-\nulldelimiterspace} {d\xi }} = \Lambda _s^{\left( 1 \right)} \left( {\left[ {0,\xi } \right)} \right)\exp \left( { - \Lambda _s \left( {\left[ {0,\xi } \right)} \right)} \right)$, since $\Pr \left\{ {L_s^{\left( 0 \right)}  < \xi } \right\} = \exp \left( { - \Lambda _s \left( {\left[ {0,\xi } \right)} \right)} \right)$. The equalities in (a) and (b) in \eqref{Eq_3__SecIV} and \eqref{Eq_4__SecIV}, respectively, follow by noting that the terms multiplying the Kronecker's delta function simplify with each other.
\subsection{Proof of Corollary \ref{Pcov_Cor}}
Since LOS and NLOS links have the same distribution, the SNR in \eqref{Eq_1__SecIV} can be equivalently re-written as ${\rm{SNR}} = {{{\mathsf{P}}G^{(0)} \left| {h^{\left( {\rm{0}} \right)} } \right|^2 } \mathord{\left/ {\vphantom {{{\mathsf{P}}G^{(0)} \left| {h^{\left( {\rm{0}} \right)} } \right|^2 } {\left( {\sigma _N^2 L^{\left( 0 \right)} } \right)}}} \right. \kern-\nulldelimiterspace} {\left( {\sigma _N^2 L^{\left( 0 \right)} } \right)}}$. Thus, the coverage can be written as ${\rm{P}}^{\left( {{\mathop{\rm cov}} } \right)} \left( {\rm{T}} \right) = \Pr \left\{ {L^{\left( 0 \right)}  < {{{\mathsf{P}}G^{(0)} \left| {h^{\left( {\rm{0}} \right)} } \right|^2 } \mathord{\left/ {\vphantom {{{\mathsf{P}}G^{(0)} \left| {h^{\left( {\rm{0}} \right)} } \right|^2 } {\left( {\sigma _N^2 {\rm{T}}} \right)}}} \right. \kern-\nulldelimiterspace} {\left( {\sigma _N^2 {\rm{T}}} \right)}}} \right\} = {\mathbb{E}}_{\left| {h^{\left( {\rm{0}} \right)} } \right|^2 } \left\{ {F_{L^{\left( 0 \right)} } \left( {{{{\mathsf{P}}G^{(0)} \left| {h^{\left( {\rm{0}} \right)} } \right|^2 } \mathord{\left/ {\vphantom {{{\rm{P}}G^{(0)} \left| {h^{\left( {\rm{0}} \right)} } \right|^2 } {\left( {\sigma _N^2 {\rm{T}}} \right)}}} \right. \kern-\nulldelimiterspace} {\left( {\sigma _N^2 {\rm{T}}} \right)}}} \right)} \right\}$. The proof follows from \textit{Lemma \ref{Intensity_Lemma}}, \textit{Lemma \ref{CDF_Lemma}} and \textit{Remark \ref{Remark_GeneralPPP}}, since $L^{\left( 0 \right)}  = \left. {\Phi ^{\left( 0 \right)} } \right|_{{\mathcal{A}}_{{\rm{LOS}}}  = {\mathcal{A}}_{{\rm{NLOS}}}  = 1}$.
\subsection{Proof of Proposition \ref{Pcov_PropApprox}}
If ${{{\mathsf{P}}G^{\left( 0 \right)} } \mathord{\left/ {\vphantom {{{\mathsf{P}}G^{\left( 0 \right)} } {\sigma _N^2 }}} \right. \kern-\nulldelimiterspace} {\sigma _N^2 }} \gg 1$, the average rate can be approximated as ${\mathop{\rm R}\nolimits}  = {\mathbb{E}}_{{\rm{SNR}}} \left\{ {{\rm{BW}}\log _2 \left( {1 + {\rm{SNR}}} \right)} \right\} \approx {\mathbb{E}}_{{\rm{SNR}}} \left\{ {{\rm{BW}}\log _2 \left( {{\rm{SNR}}} \right)} \right\}$, where the SNR is defined in \eqref{Eq_1__SecIV}. Thus, \eqref{Eq_9} can be simplified as ${\rm{R}} \approx \left( {{{{\rm{BW}}} \mathord{\left/ {\vphantom {{{\rm{BW}}} {\ln \left( 2 \right)}}} \right.  \kern-\nulldelimiterspace} {\ln \left( 2 \right)}}} \right)\int\nolimits_0^{ + \infty } {{\rm{P}}_{{\mathop{\rm cov}} } \left( {e^t } \right)dt}$. By inserting the coverage in \eqref{Eq_2__SecIV} in this approximated expression of the rate, the proof follows by swapping the order of integration and by using the following notable integral:
\begin{equation}
\label{Eq_App2__3}
{\mathcal{J}}_s \left( x \right) = \frac{1}{2}\int\nolimits_0^{ + \infty } {{\rm{erfc}}\left( {\frac{{{\rm{ln}}\left( {{{e^t x} \mathord{\left/
 {\vphantom {{e^t x} {\gamma ^{\left( 0 \right)} }}} \right.
 \kern-\nulldelimiterspace} {\gamma ^{\left( 0 \right)} }}} \right) - \mu _s }}{{\sqrt 2 \sigma _s }}} \right)dt}  = \frac{1}{2}\int\nolimits_0^{ + \infty } {{\rm{erfc}}\left( {\frac{{{\rm{ln}}\left( {{x \mathord{\left/
 {\vphantom {x {\gamma ^{\left( 0 \right)} }}} \right.
 \kern-\nulldelimiterspace} {\gamma ^{\left( 0 \right)} }}} \right) - \mu _s  + t}}{{\sqrt 2 \sigma _s }}} \right)dt}
\end{equation}
\noindent whose closed-form solution is available in \eqref{Eq_6__SecIV}.
\clearpage
\twocolumn
\end{document}